\documentclass[prd,twocolumn,superscriptaddress,floatfix,amsmath,amssymb,amsfonts,nofootinbib,longbibliography]{revtex4-2}

\usepackage{float} 
\usepackage{scalerel}
\usepackage[normalem]{ulem}
\usepackage[english]{babel}
\usepackage{graphicx}
\usepackage{dcolumn}
\usepackage{bm}
\usepackage{blindtext}
\usepackage{verbatim}
\usepackage{relsize}
\usepackage{mathrsfs}
\usepackage{musicography}
\usepackage{amsmath}
\usepackage{blindtext}
\usepackage{cancel}
\usepackage{physics}
\usepackage{epstopdf}
\usepackage{mathtools}
\usepackage{blindtext}
\usepackage{tensor}
\usepackage{color}
\usepackage[dvipsnames]{xcolor}
\usepackage[usenames,dvipsnames]{pstricks}
\usepackage{epsfig}
\usepackage{pst-grad} 
\usepackage{pst-plot} 
\usepackage{hyperref}
\usepackage{verbatim}
\usepackage{slashed}
\usepackage{dsfont}
\usepackage{tikz}
\usepackage{subfigure}

\newsavebox\mybox

\newcommand{\mf}{\mathsf}

\newcommand{\ii}{\mathrm{i}}

\newcommand{\tc}[1]{\textsc{#1}}

\newcommand{\too}[1]{\textcolor{orange}{#1}}
\newcommand{\crr}[1]{\textbf{\textcolor{red}{[[#1]]}}}

\newcommand{\brakket}[1]{\langle #1 \rangle}

\newcommand{\be}{\begin{equation}}
\newcommand{\ee}{\end{equation}}
\newcommand{\bea}{\begin{eqnarray}}
\newcommand{\eea}{\end{eqnarray}}
\definecolor{rured}{rgb}{0.745,0.192,0.102}
\definecolor{arsenic}{rgb}{0.23, 0.27, 0.29}

\newcommand{\phih}{\hat\phi}
\newcommand{\pih}{\hat\pi}
\newcommand{\rhoh}{\hat\rho}

\newcommand{\todiscuss}[1]{\color{forestgreen}}

\definecolor{mathc1}{rgb}{0.37,0.51,0.71}
\definecolor{mathc2}{rgb}{0.88,0.61,0.14}
\definecolor{mathc3}{rgb}{0.56,0.69,0.19}

\makeatletter 

\renewcommand\onecolumngrid{
\do@columngrid{one}{\@ne}%
\def\set@footnotewidth{\onecolumngrid}
\def\footnoterule{\kern-6pt\hrule width 1.5in\kern6pt}%
}

\renewcommand\twocolumngrid{
        \def\footnoterule{
        \dimen@\skip\footins\divide\dimen@\thr@@
        \kern-\dimen@\hrule width.5in\kern\dimen@}
        \do@columngrid{mlt}{\tw@}
}%
\makeatother

\allowdisplaybreaks[1] 

\begin{document}

\title{The multimode nature of spacetime entanglement in QFT}

\author{Ivan Agullo}

\affiliation{Department of Physics and Astronomy, Louisiana State University, Baton Rouge, LA 70803, USA}

\author{B\'eatrice Bonga}

\affiliation{Institute for Mathematics, Astrophysics and Particle Physics, Radboud University, 6525 AJ Nijmegen, The Netherlands}

\author{Eduardo Mart\'in-Mart\'inez}

\affiliation{Perimeter Institute for Theoretical Physics, Waterloo, Ontario, N2L 2Y5, Canada}
\affiliation{Department of Applied Mathematics, University of Waterloo, Waterloo, Ontario, N2L 3G1, Canada}
\affiliation{Institute for Quantum Computing, University of Waterloo, Waterloo, Ontario, N2L 3G1, Canada}

\author{Sergi Nadal-Gisbert}
\affiliation{QTF Centre of Excellence, Department of Physics,
University of Helsinki, FI-00014 Helsinki, Finland}

\author{T. Rick Perche}

\affiliation{Perimeter Institute for Theoretical Physics, Waterloo, Ontario, N2L 2Y5, Canada}
\affiliation{Department of Applied Mathematics, University of Waterloo, Waterloo, Ontario, N2L 3G1, Canada}
\affiliation{Institute for Quantum Computing, University of Waterloo, Waterloo, Ontario, N2L 3G1, Canada}

\author{Jos\'e Polo-G\'omez}

\affiliation{Department of Applied Mathematics, University of Waterloo, Waterloo, Ontario, N2L 3G1, Canada}
\affiliation{Institute for Quantum Computing, University of Waterloo, Waterloo, Ontario, N2L 3G1, Canada}
\affiliation{Perimeter Institute for Theoretical Physics, Waterloo, Ontario, N2L 2Y5, Canada}

\author{Patricia Ribes-Metidieri}

\affiliation{Institute for Mathematics, Astrophysics and Particle Physics, Radboud University, 6525 AJ Nijmegen, The Netherlands}

\author{Bruno de S. L. Torres}
\affiliation{Perimeter Institute for Theoretical Physics, Waterloo, Ontario, N2L 2Y5, Canada}
\affiliation{Institute for Quantum Computing, University of Waterloo, Waterloo, Ontario, N2L 3G1, Canada}
\affiliation{Department of Physics and Astronomy, University of Waterloo, Waterloo, Ontario, N2L 3G1, Canada}

\begin{abstract}

     We demonstrate the presence of multimode entanglement in the vacuum state of a free, massless scalar quantum field in four-dimensional flat spacetime between two sets of field modes, each contained within a spacetime region that is causally disconnected from the other. This is true despite the fact that entanglement between pairs of individual field modes is sparse and appears only when the two individual modes are carefully selected. Our results reveal that, while entanglement between individual modes is limited, bipartite multimode entanglement in quantum field theory is ubiquitous. We further argue that such multimode entanglement is operationally extractable, and it forms the basis of the entanglement commonly discussed in entanglement harvesting protocols.
    


\end{abstract}
\maketitle

\section{Introduction}

One of the many challenges in studying the correlations of a multipartite quantum system lies in understanding how entanglement is distributed throughout its parts. Indeed, classifying multipartite entanglement is an active research program, and efficient and general techniques for its classification are only available for sufficiently low dimensional bipartite systems~\cite{MultipartiteEntanglementReview, Watrous_2018, Gour2013, Beckey2021,BurchardtThesis,Burchardt2024}. Despite these challenges, entanglement remains one of the most relevant quantum resources, with applications to quantum communication protocols~\cite{EntanglementBasedQC}, cryptography~\cite{Ekert1991} and computational tasks~\cite{Schor}.

Quantifying entanglement is particularly elusive in the context of quantum field theory (QFT), and the task becomes specially challenging when attempting to quantify field entanglement  between two finite non-overlapping spacetime regions. Among the reasons why this is the case is the fact that the Hilbert space of a QFT does not factor as a tensor product of Hilbert spaces associated with different regions of spacetime. Thus, one cannot formally define reduced density matrices of a quantum field associated with finite spacetime regions.


One technique for quantifying the entanglement that is physically accessible by local observers is the protocol of entanglement harvesting. In essence, it consists of coupling two uncorrelated probes to a quantum field in distinct localized regions of spacetime. After the interaction, the probes end up in an entangled state by means of extracting entanglement previously present in the field, which exists even if their interaction regions are spacelike separated. This protocol was first explored in~\cite{Valentini1991,reznik1,reznik2}, and further studied  in~\cite{Pozas-Kerstjens:2015,Pozas2016}, where the protocol was also considered in physical setups, such as in the context of light-matter interaction. Since then, entanglement harvesting has been studied in a plethora of scenarios in both flat and curved spacetimes~\cite{Salton:2014jaa,Ng2014,mutualInfoBH,freefall,HarvestingSuperposed,Henderson2019,bandlimitedHarv2020,ampEntBH2020,carol,boris,ericksonWhen,threeHarvesting2022,twist2022,cisco2023harvesting,SchwarzchildHarvestingWellDone}. 

Another way of quantifying entanglement between localized spacetime regions is by analyzing the entanglement between finite sets of degrees of freedom of the field that live in different regions of spacetime. For free fields, this analysis can be conveniently done through a phase space description of canonically commuting modes localized in  each region. In a QFT, there are uncountably infinitely many modes within any finite region of space. But this inherent difficulty can be bypassed by either selecting some finite set of modes that somehow can represent enough degrees of freedom to meaningfully capture the entanglement structure of the field, or by setting the field in a lattice (where the number of modes is finite) and then studying its continuum limit.


Using this phase space description for localized modes in a quantum field theory (QFT), a recipe for finding the form of the most entangled modes respectively localized in two non-overlapping regions of space was provided in~\cite{Natalie1,NatalieUVIR,Natalie2,Natalie3}. This recipe was then applied to a lattice field theory in various spacetime dimensions. 
However, the modes which contain the entanglement between the two regions happen to have a special form. This is aligned with the results in \cite{ubiquitous}, where it was shown that randomly chosen pairs of modes localized in spacelike separated regions are typically not entangled.
The collective message from these analyses is that, while it is indeed possible to find two entangled modes in spacelike separated regions, those modes are very fine-tuned (and, in particular, not spherically symmetric).
Interestingly, when two localized quantum probes couple to the field in spacelike separated spacetime regions, the probes are able to harvest entanglement from the field~\cite{reznik1,Pozas-Kerstjens:2015}. This can be true even when the coupling between the probes and the field is determined by spatial profiles similar to those that defined the field modes in~\cite{ubiquitous}, which do not hold any pairwise entanglement. In other words, none of the individual field degrees of freedom that one probe couples to at one instant is entangled with any of the degrees of freedom that the other probe interacts with at any other instant. This poses an apparent contradiction; resolving it is the primary goal of this work. As we will demonstrate, the resolution lies in the probe's ability to harvest multimode entanglement.

This manuscript is organized as follows. In section~\ref{sec:entInQFT}, we discuss the challenges of quantifying entanglement in quantum field theory. In section~\ref{sec:modesQFT}, we describe the procedure we use to define field modes in spatial slices and briefly review the results of~\cite{ubiquitous}. Section~\ref{sec:entHarv}  reviews the protocol of entanglement harvesting, and discusses the necessary conditions for two spacelike separated probes to be able to harvest entanglement from the vacuum of a QFT.  Section~\ref{sec:paradox} discusses in some detail the apparent tension between the sparse nature of pair-wise entanglement in QFT and the ability of local probes to harvest entanglement. 
In section~\ref{sec:results}
 we quantify multimode entanglement in quantum field theory and compare our findings with the protocol of entanglement harvesting. This section contains the main results of this article. Finally, we summarize our results and put them in a broader perspective in section~\ref{sec:discussions}.

 As a note for readers in a hurry, sections \ref{sec:entInQFT} to \ref{sec:entHarv} primarily contain review material. These sections aim to make this article self-contained for a broad audience by providing background information on the tools and concepts we use. Readers already familiar with these topics may wish to skip ahead to section \ref{sec:paradox}.


\section{The challenges of quantifying entanglement in QFT}\label{sec:entInQFT}
    In this section we provide some background on the problem of characterizing entanglement in QFT. We begin with a quick review of the quantum theory of a real Klein-Gordon field, which is the main example we will focus on in this work. We then briefly recap how one usually thinks about entanglement in quantum mechanics, and highlight what makes QFTs qualitatively different. This background will be important to contextualize our work within the broader literature on entanglement in field theory, and also help to motivate the approach that we will adopt in order to tackle this problem in the rest of the paper.

\subsection{Quick review of the quantum theory of a free real scalar field}\label{sub:KGfieldreview}
    
    Let $(\mathcal{M}, g_{ab})$ be a globally hyperbolic spacetime, where $\mathcal{M}$ is a differentiable manifold and $g_{ab}$ is a Lorentzian metric on $\mathcal{M}$. In the algebraic approach to QFT, the starting point of a quantum field theory is an \emph{algebra of observables} $\mathcal{A}(\mathcal{M})$ on $\mathcal{M}$, together with a collection of subalgebras $\mathcal{A}(R)$ associated to causally complete spacetime regions \mbox{$R\subset \mathcal{M}$}. Each local algebra is assumed to be what we call a \emph{unital $\ast$-algebra}, which just means that the algebra possesses an identity element $\hat{\mathds{1}}$, as well as an involution with the properties of the adjoint.\footnote{The name ``$\ast$-algebra'' is inherited from the fact that mathematicians customarily denote the adjoint of a given element $\hat{A}\in\mathcal{A}$ by $\hat{A}^\ast$. In this paper, however, we will stick to the physicist's convention, where the adjoint of $\hat{A}$ is denoted by $\hat{A}^\dagger$.} The intuitive idea is that, for a given spacetime subregion $R$, the algebra $\mathcal{A}(R)$ comprises the representation of the field degrees of freedom that are supported in $R$. In the case of a scalar field, this local algebra of observables is generated by the identity element $\hat{\mathds{1}}$ together with smeared field operators of the form
    \begin{equation}\label{eq:spacetimesmearedfield}
        \hat{\phi}(f) = \int_{\mathcal{M}} \dd V\,\hat{\phi}(\mf x)f(\mf x),
    \end{equation}
    where $\dd V$ is the volume form on spacetime, and \mbox{$f(\mf x) \in C_{0}^{\infty}(\mathcal{M})$}, with $C_{0}^{\infty}(\mathcal{M})$ denoting the set of smooth, compactly supported functions on $\mathcal{M}$. The quantum field $\hat{\phi}(\mf x)$ in this context is thus understood as an operator-valued distribution (with $\hat{\phi}(f)$ being the output of the distribution when acting on the test function $f(\mf x)$), and the algebra $\mathcal{A}(R)$ is generated by smeared field operators $\hat{\phi}(f)$ where the support of the function $f(\mf x)$ is contained in $R$. 

    We then impose additional conditions on the algebra of observables associated to each local region of spacetime, in order for the collection of algebras $\mathcal{A}(R)$ to define a bonna-fide relativistic QFT on $\mathcal{M}$. For a real Klein-Gordon field, these conditions are that the field is Hermitian---i.e., $\hat{\phi}(f^*) = \hat{\phi}(f)^\dagger$ for any (possibly complex) test function $f$---, and that it satisfies the commutation relations
    \begin{equation}\label{eq:covariantcommutationrelations}
        \comm{\hat{\phi}(f)}{\hat{\phi}(g)} = \ii E(f, g)\hat{\mathds{1}},\,\, \forall f, g \in C^{\infty}_{0}(\mathcal{M}),
    \end{equation}
    and the Klein-Gordon equation of motion---which, since $\hat{\phi}$ is a distribution, is more appropriately stated as
    \begin{equation}\label{Eq: kernelphi}
    \begin{gathered}
        \hat{\phi}(h) = 0 \,\,\text{whenever} \,\,h(\mf x)\,\, \text{is of the form} \\ h = (g^{ab}\nabla_a\nabla_b - m^2 - \xi R)f, \,\,\,\,f\in C^\infty_{0}(\mathcal{M}),
    \end{gathered}
    \end{equation}
    where $\xi\geq0$ is a dimensionless constant, and $R$ is the Ricci scalar of the background. The object $E(f, g)$ in Eq.~\eqref{eq:covariantcommutationrelations} is given by
    \begin{equation}
        E(f, g) = \int \dd V\,\dd V' E(\mf x, \mf x')f(\mf x)g(\mf x'),
    \end{equation} 
    where $E(\mf x, \mf x')$ is the \emph{causal propagator}---i.e., the difference between retarded and advanced Green's functions---for the Klein-Gordon equation in $\mathcal{M}$. Equation~\eqref{eq:covariantcommutationrelations} is equivalent to the usual equal-time canonical commutation relations between field and conjugate momentum (which would be the more common starting point in canonical quantization), but restated in a spacetime-covariant way that does not require the specification of any notion of ``equal time''. Also note that, since the causal propagator $E(\mf x, \mf x')$ vanishes if the points $\mf x$ and $\mf x'$ are spacelike separated, the algebra~\eqref{eq:covariantcommutationrelations} guarantees that observables associated to spacelike separated regions commute. This is a condition known as \emph{microcausality}, which is a property that every relativistic QFT is expected to satisfy.
    
    It will be useful to note that, thanks to the Klein-Gordon equation of motion, Eq.~\eqref{eq:spacetimesmearedfield} can equivalently be written as
    \begin{equation}\label{eq:symplecticsmearing}
        \hat{\phi}(f) = \int_{\Sigma} \dd\Sigma\,n^a\left(\hat{\phi}\,\nabla_a F - F\,\nabla_a\hat{\phi}\right),
    \end{equation}
    where $\Sigma$ is a spacelike Cauchy surface for $\mathcal{M}$, $\dd\Sigma$ is the induced volume form on $\Sigma$, $n^a$ is the future-oriented unit normal to $\Sigma$, and $F(\mf x)$ is a function defined by
    \begin{equation}\label{Eq: smearing to Cauchy}
        F(\mf x) \coloneqq (Ef)(\mf x) = \int_{\mathcal{M}} \dd V' E(\mf x, \mf x')f(\mf x').
    \end{equation}
   This expresses the familiar fact that, whenever the Klein-Gordon equation in $\mathcal{M}$ admits a well-posed initial value formulation (i.e., whenever the spacetime in question is globally hyperbolic), $\mathcal{A}(\mathcal{M})$ can be constructed out of the field and its conjugate momentum at a constant time slice---where the notion of ``surface of constant time" is given by the Cauchy surface $\Sigma$, and $n^a\nabla_a\hat{\phi}$ plays the role of the conjugate momentum.

   After giving the general properties of the algebra of observables, we complete the definition of a QFT by providing a notion of \emph{states}. In the most abstract sense, a state on an algebra $\mathcal{A}$ (or, more concisely, an algebraic state) is simply a linear map $\omega: \mathcal{A}\to\mathbb{C}$ from elements of the algebra to the complex numbers, with the property that $\omega(\hat{\mathds{1}}) = 1$ (i.e., $\omega$ maps the identity element of the algebra to the number $1$) and $\omega(\hat{A}^\dagger\hat{A}) \geq 0$ for any element $\hat{A}\in \mathcal{A}$ (i.e., $\omega$ maps positive operators in the algebra to nonnegative numbers). The idea is that, for any given element $\hat{A}\in\mathcal{A}$, the output $\omega(\hat{A})$ corresponds to the expectation value of the observable $\hat{A}$ in the state $\omega$. In the algebraic sense, a state is said to be \emph{mixed} if it can be expressed as a convex linear combination of two distinct states---i.e., if there are states $\omega_1$ and $\omega_2$ and $\alpha\in(0, 1)$ such that $\omega = (1-\alpha)\omega_1 + \alpha\omega_2$, with $\omega_1\neq\omega_2$. An algebraic state is called \emph{pure} if it is not mixed.

   Physical requirements can impose some further constraints on the map that defines an algebraic state. For $\omega$ to be a good candidate for a physical state of a QFT, we expect, at the very least, that physically relevant observables have finite expectation values, possibly after some renormalization scheme is implemented. It turns out that this requirement is quite restrictive. For example, consider the state of the field is such that its two-point function $\omega(\hat{\phi}(\mf x)\hat{\phi}(\mf y))$ takes on the so-called Hadamard form
    \begin{equation}\label{eq:hadamardcondition}
       \omega(\hat{\phi}(\mf x)\hat{\phi}(\mf y)) = \dfrac{U(\mf x, \mf y)}{\sigma(\mf x, \mf y)} + V(\mf x, \mf y)\log\left[ \sigma(\mf x, \mf y)\right] + H_{\omega}(\mf x, \mf y),
   \end{equation}
    where $\sigma(\mf x, \mf y)$ is Synge's world function (i.e., half of the squared geodesic distance between the spacetime points $\mf x$ and $\mf y$), $U$ and $V$ are two smooth functions that are fully fixed by the background metric $g_{ab}$ and the Klein-Gordon equation, and $H_\omega$ is another smooth function that encodes all the possible state dependence~\cite{WaldQFT}.\footnote{Note that, just like $\hat{\phi}(\mf x)$ is not really an operator but an operator-valued distribution, the ``two-point function'' $\omega(\hat{\phi}(\mf x)\hat{\phi}(\mf y))$ is not actually a function but a bi-distribution, which should be smeared against test functions on $\mathcal{M}\times\mathcal{M}$ to produce truly observable quantities.} A state of this form behaves locally as the Minkowski vacuum as the two points $\mf{x}$ and $\mf{y}$ get closer to each other, and the stress-energy density is guaranteed to be renormalizable using local and covariant methods.

For background spacetimes with high symmetry, a privileged choice of algebraic state can be made by requiring that expectation values behave nicely under the action of the symmetries of the background. This can be used, in particular, to specify a preferred notion of \emph{vacuum state} for the QFT. In Minkowski spacetime, for example, requiring the vacuum to be invariant under Poincaré symmetry and scale transformations (for massless fields) fixes the functional form of the vacuum two-point function to be of Hadamard form (cf. Eq.~\eqref{eq:hadamardcondition}), with $U = \text{constant}$, and $V = H_\omega = 0$, where the geodesic distance between $\mf x$ and $\mf y$ is simply given by  $(x-y)^2$. Note also that, since a Klein-Gordon field is free, the ground state is guaranteed to be Gaussian. Therefore, specifying the two-point function fully determines the state, with higher-order correlation functions being expressible in terms of it via Wick contractions.

In curved spacetimes, the $1/\sigma$ term in~\eqref{eq:hadamardcondition} dominates the behavior of the two-point function of any Hadamard state in the coincidence limit (that is, when $\mf y \to \mf x$), regardless of the background curvature. Moreover, when both $\mf x$ and $\mf y$ are contained in a sufficiently small convex normal neighbourhood of some point $\mf p$, Synge's world function $\sigma(\mf x, \mf y)$ approximates its ``flat-spacetime'' expression $\eta_{\mu\nu}(x^\mu - y^\mu)(x^\nu - y^\nu)/2$, in Riemann normal coordinates at $\mf p$. One can therefore see that a direct consequence of the Hadamard condition~\eqref{eq:hadamardcondition} is that the two-point function of any ``finite-energy'' state of a QFT (that is, a state such that the energy-momentum tensor admits a finite renormalized expectation value) must have the same singularity structure at short distances as the vacuum state in Minkowski space, with the correlation function diverging polynomially as the distance between the two points approaches zero.

The takeaway messages of this section are: 1) the degrees of freedom of a QFT in a given region of spacetime are described in terms of an algebra of observables associated to that region; 2) states in a QFT are maps that connect elements of the algebra to their expectation values; and 3) all physically acceptable states (in particular, states with finite energy and momentum) display the same universal short-distance behavior for the two-point function of the field as the two points approach each other. These facts will turn out to be essential to understand why entanglement is tricky to quantify in QFT, as we explain in the next section.

\subsection{Why analyzing entanglement in QFT is hard}\label{sub:entQFThard}

The basic introduction to the concept of entanglement in quantum mechanics usually goes as follows. Consider two physical systems $A$ and $B$, with respective Hilbert spaces $\mathcal{H}_A$ and $\mathcal{H}_B$. The Hilbert space of the joint system $AB$ is thus defined as the tensor product \mbox{$\mathcal{H} = \mathcal{H}_{A}\otimes \mathcal{H}_{B}$}. Having this decomposition of the joint Hilbert space in mind, a pure quantum state $\ket{\psi}\in\mathcal{H}$ is said to be \emph{entangled} if it is not \emph{separable}---i.e., if it cannot be written as $\ket{\psi} = \ket{\chi}\otimes\ket{\phi}$ for some $\ket{\chi}\in\mathcal{H}_A$, $\ket{\phi} \in \mathcal{H}_B$. The entanglement between subsystems $A$ and $B$ can then be directly quantified by the \emph{entanglement entropy}, which is given by the von Neumann entropy of the reduced state of either $A$ or $B$ (the von Neumann entropies of both reduced density matrices $\hat{\rho}_A = \Tr_B\left(\ket{\psi}\bra{\psi}\right)$ and $\hat{\rho}_B = \Tr_A\left(\ket{\psi}\bra{\psi}\right)$ in this case are guaranteed to be the same since the joint state of $AB$ is pure).

It is in general also important to have a notion of entanglement when the joint state of the quantum system $AB$ is not pure. This is relevant, for example, when $AB$ is itself a subsystem of a larger system $ABC$, one of whose components (here denoted by $C$) can be entangled with $AB$, thus resulting in a mixed state $\hat{\rho}_{AB}$ for the quantum system of interest. Entanglement in mixed states is again defined as lack of separability, but the notion of ``separability'' is slightly modified: now, a mixed state is considered separable not only if it can be expressed as a product $\hat{\rho}_A\otimes\hat{\rho}_B$ for a pair of states $\hat{\rho}_A$ and $\hat{\rho}_B$, but also if it can be expressed as $\sum_i p_i \hat{\rho}_A^{(i)}\otimes\hat{\rho}_B^{(i)}$ for some set of probabilities $\{p_i\}$ and an ensemble of states $\{\hat{\rho}_A^{(i)}\}$ and $\{\hat{\rho}_B^{(i)}\}$.  This definition allows states to be called separable while still containing classical correlations, which is something we did not have to worry about in the pure state case because all correlations in a pure state are quantum in nature. 

Characterizing entanglement in full generality for mixed states is notoriously difficult. In fact, deciding whether a given (mixed) bipartite quantum state is entangled or separable is an NP-hard problem in general~\cite{entanglementNPHard}. Fortunately, there exist simple quantities that one can easily compute for any mixed state $\hat{\rho}_{AB}$ which turn out to be quite useful in the study of mixed-state entanglement. One of these quantities is the \emph{negativity}, defined as
\begin{equation}\label{eq:negativity}
    \mathcal{N}(\hat{\rho}_{AB}) \coloneqq \dfrac{||\hat{\rho}_{AB}^{T_B}||_1 - 1}{2},
\end{equation}
where $||\hat{O}||_1 \equiv \Tr\sqrt{\hat{O}^\dagger\hat{O}}$ is the $1$-norm of the operator $\hat{O}$, and $\hat{\rho}_{AB}^{T_B}$ is the partial transpose of the density matrix $\hat{\rho}_{AB}$ with respect to subsystem $B$. 
By noting that $\Tr\hat{\rho}_{AB}^{T_B} = \Tr\hat{\rho}_{AB} = 1$, the definition~\eqref{eq:negativity} can simply be restated as the magnitude of the sum of all negative eigenvalues of $\hat{\rho}_{AB}^{T_B}$. 

It is clear that the partial transpose of any separable mixed state will still be a  density matrix, and thus  $\Tr\hat{\rho}_{AB}^{T_B}$ equals one and the negativity vanishes. 
Therefore, nonzero negativity guarantees that the state is entangled. This gives us a very efficient way of detecting entanglement, since the spectrum of the partially transposed density matrix is very easily computable (something that cannot be said of virtually any other entanglement measure for mixed states). Beyond this use to identify entangled states, the negativity is also an entanglement \emph{monotone}, meaning that it can never increase under local operations and classical communication (LOCC)~\cite{VidalNegativity}, which makes it a potential candidate to also `quantify' the amount of entanglement in at least some families of states. Finally, a closely related quantity to the negativity, called the \emph{logarithmic negativity}
\begin{equation}
    E_{\mathcal{N}}(\hat{\rho}_{AB}) \coloneqq \log(2\mathcal{N}(\hat{\rho}_{AB})+1) = \log_2 ||\hat{\rho}_{AB}^{T_B}||_1\, ,
\end{equation}
actually provides an upper bound on another important entanglement measure called the \emph{distillable entanglement}, which roughly characterizes how much ``pure entanglement'' can be extracted from the state $\hat{\rho}_{AB}$ by using only local operations and classical communication.\footnote{In slightly more precise words, the distillable entanglement quantifies the number of EPR pairs per copy of $\hat{\rho}_{AB}$ that can be extracted from the state $\hat{\rho}_{AB}^{\otimes N}$ by using only local operations and classical communication in the limit $N\to \infty$. For more details, see~\cite{entanglementmeasuresreview}.} In particular, states with zero negativity---also called positive-partial-transpose  (or PPT) states---also have zero distillable entanglement~\cite{VidalNegativity}. All of these facts make the negativity a very useful quantity to study in the context of entanglement in mixed states.

{\color{black}One can now ask how to adapt this general characterization of entanglement to the case of QFTs. The algebraic approach to QFT that we described in Subsection~\ref{sub:KGfieldreview} did not start with a Hilbert space, but the recipe we laid out has enough ingredients to accommodate it. This is thanks to the \emph{Gelfand-Naimark-Segal (GNS) representation theorem}, which states the following: for any unital $\ast$-algebra $\mathcal{A}$ and given an algebraic state $\omega$ on $\mathcal{A}$, there always exists a Hilbert space $\mathcal{H}_\omega$ and a vector $\ket{\Omega_\omega}\in\mathcal{H}_\omega$, together with a representation $\pi_\omega$ mapping elements of $\mathcal{A}$ to (possibly unbounded) operators on $\mathcal{H}_\omega$, such that the set $\pi_\omega(\mathcal{A})\ket{\Omega_\omega}$ is dense in $\mathcal{H}_\omega$, and one can represent the action of the state $\omega$ on any element $\hat{A}\in\mathcal{A}$ as
\begin{equation}
    \omega(\hat{A}) = \bra{\Omega_\omega}\pi_\omega(\hat{A})\ket{\Omega_\omega}.
\end{equation}
   This representation is, moreover, unique up to unitary equivalence. Note that, in the GNS representation, the algebraic state $\omega$ is always represented as a vector $\ket{\Omega_\omega}$, regardless of whether $\omega$ was mixed or pure. This means that, when applied to regular quantum mechanical systems where we would expect to describe a mixed state on a system with Hilbert space $\mathcal{H}$ as a density matrix $\hat{\rho}$, the GNS representation instead expresses $\hat{\rho}$ as its canonical purification $\ket{\sqrt{\rho}}$, which is a vector in the doubled Hilbert space $\mathcal{H}\otimes\mathcal{H}$.\footnote{A very common example of this is when $\hat{\rho}$ is thermal, in which case the GNS representation leads to the so-called thermofield double state.} In the algebraic language, where we do not have an \emph{a priori} independent definition of a Hilbert space, the difference between mixed and pure states in the algebraic sense amounts to a difference in properties of the representation $\pi_{\omega}$. Namely, if $\omega$ is mixed in the algebraic sense (i.e., if it can be written as a nontrivial convex linear combination of two distinct states on the algebra $\mathcal{A}$), then $\pi_\omega$ gives a \emph{reducible} representation of $\mathcal{A}$, whereas if $\omega$ is pure, the representation is irreducible.
   
   In our case of interest, we can take the algebra to be the algebra of smeared field operators satisfying canonical commutation relations, and the state $\omega$ to be a Hadamard state $\ket{\Omega}$ of the Klein-Gordon field. In this case, the Hilbert space $\mathcal{H}$ obtained through the GNS representation simply corresponds to the Fock space from canonical quantization, and we recover the more usual picture from standard textbook introductions to QFT.}
   
Now, given a spatial subregion $\Sigma_A\subset \Sigma$ of some Cauchy surface $\Sigma$, one can talk about the subset of field observables that are ``supported in $\Sigma_A$'' by restricting to field operators $\hat{\phi}(f)$ such that both $F\big|_{\Sigma}$ and $n^a\nabla_a F\big|_{\Sigma}$ appearing in Eq.~\eqref{eq:symplecticsmearing} are only nonvanishing in $\Sigma_A$. 
It would perhaps be natural to think that the global Hilbert space $\mathcal{H}$ of the QFT could be decomposed as something of the form $\mathcal{H}_{A}\otimes\mathcal{H}_{\bar{A}}$, where $\mathcal{H}_{A}$ and $\mathcal{H}_{\bar{A}}$ are local Hilbert spaces associated with the spatial regions $\Sigma_A$ and $\Sigma_{\bar{A}} = \Sigma\setminus \Sigma_A$, respectively; and if that were the case, one would in principle be able to talk about entanglement measures in QFT by using the same tools employed in the nonrelativistic setting. For instance, the entanglement between the two complementary regions $\Sigma_A$ and $\Sigma_{\bar{A}}$ in the vacuum of the QFT could be obtained by directly computing the von Neumann entropy of the reduced density matrix $\hat{\rho}_A = \Tr_{\bar{A}}\left(\ket{\Omega}\!\bra{\Omega}\right)$ associated with the spatial subregion $\Sigma_A$. Similarly, if we considered two arbitrary spatial subregions $\Sigma_A$ and $\Sigma_B$, we could try to define a density matrix $\hat{\rho}_{AB}$ for the joint system by tracing out the degrees of freedom associated to $\Sigma\setminus(\Sigma_A\cup\Sigma_B)$, and then compute the negativity of that density matrix as a way to quantify the portion of entanglement shared between the regions $\Sigma_A$ and $\Sigma_B$ which could be distilled by agents with access to the two regions.

It turns out that this intuitive picture does not quite work: rigorously speaking, the Hilbert space of a QFT does \emph{not} factorize as a tensor product of local Hilbert spaces associated with complementary spatial regions, and it is not even technically right to assign a density matrix to the restriction of the vacuum state $\ket{\Omega}$ (or any physical state of the QFT, for that matter) to any local subregion~\cite{haag, witten}. The mathematical reason for this is that the local algebra of observables on a finite region of a relativistic QFT is what is known as a \emph{type III von Neumann algebra}, which does not admit any irreducible representation as an algebra of operators on a Hilbert space, and does not contain any nontrivial faithful operation with the properties of a ``trace''~\cite{SorceReview}. This means, in particular, that operations such as taking a partial trace over a subregion are not available, von Neumann entropies of the reduced state of a QFT on a given region are not well-defined, and one cannot even talk about the reduced state of the QFT on a local subregion in terms of a density matrix. 

There are very clear indications, however, that any regular state of a QFT must be highly entangled. For instance, if we think of a QFT as the continuum limit of a theory defined on a lattice with some minimum lattice spacing $\varepsilon$ (which does admit density matrices and entropies for subregions for any $\varepsilon>0$), we find that the entanglement entropy of the vacuum is UV divergent: more specifically, the entanglement entropy between a spatial subregion $\Sigma_A$ and its complement $\Sigma_{\bar{A}}$ typically scales as $\text{Area}(\partial\Sigma_A)/\varepsilon^{d-2}$, where $\partial\Sigma_A$ (often referred to as the \emph{entangling surface}) is the boundary separating $\Sigma_A$ and its complement, and $d$ is the number of spacetime dimensions. This UV divergence can be understood as ultimately coming from the correlations between the field degrees of freedom at arbitrarily short distances near the entangling surface. Since we know that any regular state of the QFT should display the same short-distance behavior as the vacuum, we conclude that the entanglement between complementary regions of a QFT should be formally infinite in every state in the Hilbert space of the theory.\footnote{Incidentally, this also provides a rough explanation as to why the QFT Hilbert space cannot be factorized as $\mathcal{H}_A\otimes\mathcal{H}_{\bar{A}}$. After all, if it could, then there would be states in $\mathcal{H}$ with no correlations between a subregion and its complement---but as we have just argued, this cannot happen in any finite-energy state of the QFT.} 

Another---much more sophisticated---manifestation of the high amount of entanglement in the vacuum of a QFT is illustrated in the \emph{Reeh-Schlieder theorem}. This theorem states that, for \emph{any} local algebra $\mathcal{A}(R)$ associated to a local subregion of the QFT, the entire Hilbert space of the QFT can be arbitrarily well-approximated by states of the form $\hat{A}\ket{\Omega}$ even if we restrict the operator $\hat{A}$ to be solely in $\mathcal{A}(R)$, no matter how small the region $R$ is\footnote{In slightly more precise terms, the set $\mathcal{A}(R)\ket{\Omega}$ is \emph{dense} in $\mathcal{H}$, for any local algebra $\mathcal{A}(R)$.}. If we wanted to forcibly interpret $\ket{\Omega}$ as a state in a factorizable Hilbert space $\mathcal{H}_A\otimes\mathcal{H}_{\bar{A}}$, this would necessarily mean that $\ket{\Omega}$ is ``fully entangled''---i.e., the reduced state on both $A$ and $\bar{A}$ is a density matrix with maximum rank. 

All of the technical obstructions to quantifying entanglement in QFT ultimately boil down to the fact that we are dealing with a system that contains infinitely many degrees of freedom. It is therefore natural to wonder if one could try to make sense of this ``formally infinite'' amount of entanglement in QFT in a more operational way, by somehow translating all the relevant physics of the theory to subsystems with finitely many degrees of freedom. Our goal in the rest of the paper will be to explore two strategies that attempt to do just that.

We will first show how to define a finite-dimensional subsystem of the field theory by constructing a finite set of modes of the field by smearing the field operator and its conjugate momentum in space. Due to the finite dimensionality, the associated subalgebra is of {\em type I}, meaning that this subalgebra is a factor of the total algebra and operations like partial trace onto the subsystem are well-defined. Therefore, standard tools used for finite-dimensional quantum systems to quantify entanglement, such as negativity, are applicable. Next, we will review the protocol known in the literature as \emph{entanglement harvesting}, which allows two localized probes initially in a separable state to become entangled through local interactions with a quantum field, even when the coupling region of one probe is spacelike separated from the coupling region of the other. Our ultimate goal is to connect these two notions of entanglement. By doing so, we aim to shed light on the field degrees of freedom that are relevant for entanglement harvesting and reconcile the results of entanglement harvesting with the apparent obstructions recently found in discussions of entanglement in QFT~\cite{ubiquitous}.

\section{Mode-wise analysis of entanglement in QFT}\label{sec:modesQFT}

A way to bypass the difficulties posed by the infinitely many degrees of freedom held by a quantum field in the study of its entanglement structure is to focus on just a finite number of  degrees of freedom. In this section we review how to single out a finite number of independent modes of a free scalar quantum field, and how to describe them using a phase-space formulation. We then restrict ourselves to Gaussian states of the field, for which the Gaussian formalism can be employed, and the quantification of entanglement is particularly simple. We will finish by reviewing the results of~\cite{ubiquitous}, where the entanglement between different modes of a real scalar quantum field in the vacuum state of a $(1+D)$-dimensional Minkowski spacetime was analyzed.     

\subsection{Construction of field modes}\label{Subsection: Construction of field modes}

Given a set of smooth compactly supported real functions \mbox{$\{f_1,g_1,\hdots,f_N,g_N\}\subset\mathcal{C}_0^\infty(\mathcal{M})$} such that
\begin{align}
    E(f_j,f_k) & = E(g_j,g_k) = 0, \\
    E(f_j,g_k) & = \delta_{jk},
\end{align}
 we have, using Eq.~\eqref{eq:covariantcommutationrelations}, that their associated smeared field operators satisfy
\begin{align}
[\phih(f_j),\phih(f_k)] & = [\phih(g_j),\phih(g_j)] = 0, \\
[\phih(f_j),\phih(g_k)] & = \ii \delta_{jk} \hat{\openone}.
\end{align}
That is, the two sets of $N$ operators $\{\phih(f_1),\hdots,\phih(f_N)\}$ and $\{\phih(g_1),\hdots,\phih(g_N)\}$ 
satisfy the canonical commutation relations (CCR), and therefore define a quantum system of $N$ (bosonic) degrees of freedom. Each degree of freedom is identified by {the subalgebra generated by} the canonically conjugate pair $(\phih(f_j),\phih(g_j))$, and we refer to it as a \textit{mode} of the field. {This subalgebra is isomorphic to the familiar algebra of a single harmonic oscillator.}

Ideally, one would like different choices of smearing functions $f_j, g_j$ to yield different (albeit not necessarily independent) modes of the field. However, the construction above does not have that property: indeed, the operator-valued distribution $\phih$ has a non-trivial kernel, described in Eq.~\eqref{Eq: kernelphi}, and therefore two functions whose difference lies in the kernel of $\phih$ would yield the same smeared operator, i.e., the representation \mbox{$\mathcal{C}_0^\infty(\mathcal{M}) \ni f \mapsto \phih(f)$} is not faithful. One way to explicitly quotient the space of test functions over the kernel of $\phih$ is to consider a fixed spacelike Cauchy hypersurface $\Sigma \subset \mathcal{M}$ and use the well-posedness of the initial-value problem for the Klein-Gordon equation, guaranteed by the global hyperbolicity of $\mathcal{M}$, as reviewed in Sec.~\ref{sec:entInQFT}. Recalling Eq.~\eqref{eq:symplecticsmearing}, 
\begin{align}
        \hat{\phi}(f) & = \int_{\Sigma} \dd\Sigma\,n^a\left(\nabla_a F \,\hat{\phi}- F\,\nabla_a\hat{\phi}\right) \\
        & \eqqcolon \hat O (F,n^a\nabla_a F), \nonumber
\end{align}
with $F=Ef$ given by Eq.~\eqref{Eq: smearing to Cauchy}, which is a solution of the Klein-Gordon equation. In general, given $F,G \in \mathcal{C}_0^\infty(\Sigma)$, we can consider operators of the form
\begin{equation}\label{Eq: F,G field operator}
    \hat O (F,G) =   \int_{\Sigma} \dd\Sigma\,\left(G \, \hat{\phi} - F\,n^a\nabla_a\hat{\phi}\right),
\end{equation}
so that
\begin{equation}\label{Eq: commutation F,G operators}
    [\hat O(F,G),\hat O (F',G')] = \ii \int_\Sigma \dd\Sigma \,(F G' - G F') \hat{\openone}.
\end{equation}
Reciprocally, given any pair $(F,G)$, there exists a {unique} solution $\varphi$ of the (homogeneous) Klein-Gordon equation such that $\varphi|_\Sigma = F$ and $n^a \nabla_a\varphi|_\Sigma = G$. By the properties of the causal propagator $E$, there exists $f \in \mathcal{C}_0^\infty(\mathcal{M})$ such that $\varphi = E f$, hence $\phih(f) = \hat O (F,G)$. 

In sum, because we are working with a free field theory, it is possible to represent any smeared field operator $\phih(f)$  as an operator of the form $\hat O (F,G)$, and viceversa, and unlike the map $f \mapsto \phih(f)$, the map $(F,G)\mapsto \hat O (F,G)$ is faithful, since the kernels of $E$ and $\phih$ coincide.

It is worth remarking that the previous construction can be performed with any choice of spacelike Cauchy hypersurface $\Sigma$. In fact, given two such hypersurfaces, $\Sigma_1$ and $\Sigma_2$, it is straightforward to map one representation, $\hat O_1$, into the other, $\hat O_2$: for any pair $(F_1,G_1)$, it suffices to find a homogeneous solution $\varphi$ such that $\varphi|_{\Sigma_1} = F_1$ and $n_1^a \nabla_a \varphi|_{\Sigma_1} = G_1$. Taking $F_2 = \varphi|_{\Sigma_2}$ and $G_2 = n_2^a \nabla_a \varphi|_{\Sigma_2}$, we find that $\hat O_1 (F_1,G_1) = \hat O_2 (F_2, G_2)$. 
One particular consequence of this fact is that, if we pick a foliation $\{\Sigma_t\}$ of $\mathcal{M}$, for some global time function $t$, then the operator
\begin{equation}\label{Eq: O varphi}
\!\!\!\hat O(\varphi) \coloneqq \!\!\int_{\Sigma_t}\!\!\!\! \dd\bm x\! \left(\sqrt{\mathsf{h}}\nabla_{\mathsf n}\varphi(t,\bm x) \, \hat{\phi}(t,\bm x) - \varphi(t,\bm x)\,\pih(t,\bm x)\right)\!
\end{equation}
does not depend on $t$,\footnote{\label{Heisenbergev} Notice that this expression is different from the Heisenberg evolution of the operator $\hat O(\varphi)$ from $t_0$ to $t$, which can be written as
\begin{equation}\hat O(\varphi,t) \coloneqq \!\!\int_{\Sigma_t}\!\!\!\! \dd\bm x\! \left(\sqrt{\mathsf{h}}\nabla_{\mathsf n}\varphi(t_0,\bm x) \, \hat{\phi}(t,\bm x) - \varphi(t_0,\bm x)\,\pih(t,\bm x)\right)\!.
\end{equation}
This expression differs from \eqref{Eq: O varphi} in that  the solution $\varphi$ has been ``frozen'' at the instant $t_0$.} for any given solution $\varphi$ of the Klein-Gordon equation, where $\mathsf{h}$ is the determinant of the induced metric on $\Sigma_t$, and $\pih \coloneqq \sqrt{\mathsf{h}} \nabla_{\mathsf n}\phih = \sqrt{\mathsf{h}} n^a \nabla_a\phih $ is the conjugate momentum of $\phih$ on $\Sigma_t$, with $n_a \propto (\dd t)_a$ so that $n^a n_a = -1$.  For a given $t$, we can build field and momentum operators in the following form:
\begin{align}\label{Eq: field and momentum operators}
    \hat\Phi(t,G) & \coloneqq \int_{\Sigma_t} \dd\bm x\, \sqrt{\mathsf{h}}\, G(\bm x) \phih(t,\bm x), \\
    \hat\Pi(t,F) & \coloneqq \int_{\Sigma_t} \dd\bm x\, F(\bm x) \pih(t,\bm x). 
\end{align}
which `separate' field amplitude and conjugate momentum only for a particular time $t$. In general, for some other $t'$, they will have the form of $\hat O(\varphi)$ in Eq.~\eqref{Eq: O varphi}, for the homogeneous solution $\varphi$ that satisfies the initial conditions $\varphi|_{\Sigma_t} = 0$, $\nabla_{\mf n} \varphi|_{\Sigma_t} = G(\bm x)$, for the field operator, and  $\varphi|_{\Sigma_t} = -F(\bm x)$, $\nabla_{\mf n} \varphi|_{\Sigma_t} = 0$, for the momentum operator. 

\subsection{Phase-space formulation}\label{Subsection: Phase-space formulation}

Once we have singled out a finite number $N$ of degrees of freedom,\footnote{Algebraically, this means that we went from a type III to a type I von Neumann algebra.} the resulting continuous-variable system can be naturally endowed with a symplectic structure. Specifically, we can establish a correspondence between 
linear combinations of the operators that define the $N$ degrees of freedom, and the elements of the classical phase space $\mathbb{R}^{2N}$, by  
\begin{equation}
\hat\Xi(\bm \xi) = \Omega_{\alpha\beta}\xi^\beta\hat\Xi^\alpha,
\end{equation}
where $\Omega_{\alpha\beta}$ are the components of the symplectic matrix
\begin{equation}
\bm\Omega = \bigoplus_{j=1}^N \begin{pmatrix}
0 & -1 \\
1 & 0
\end{pmatrix},
\end{equation}
$\xi^\beta$ are the canonical components of a phase-space vector $\bm\xi\in\mathbb{R}^{2N}$, and $\hat\Xi^\alpha$ are the components of a vector comprising all the canonically conjugate observables that define the $N$ modes of the system,
\begin{equation}
    \bm{\hat\Xi} = (\phih(f_1),\phih(g_1),\hdots,\phih(f_N),\phih(g_N))^\intercal.
\end{equation}
The CCR can thus be rewritten as
\begin{equation}
    [\hat\Xi^\alpha,\hat\Xi^\beta] = \ii \Omega^{\alpha\beta}\hat{\openone},
\end{equation}
where $\Omega^{\alpha\beta}$ are the components of the inverse symplectic matrix $\Omega^{-1}$, yielding the general commutation relation 
\begin{equation}
    [\hat\Xi(\xi_1),\hat\Xi(\xi_2)] = - \ii \Omega(\xi_1,\xi_2) \hat{\openone} = - \ii \Omega_{\alpha\beta}\xi_1^\alpha \xi_2^\beta \hat{\openone}.
\end{equation}

\subsection{Gaussian states}\label{Subsection: Gaussian states}

The state of a quantum field is called \textit{Gaussian} if it can be completely characterized by its one-point functions $\langle \phih(\mathsf{x}) \rangle$, and two-point functions $\langle \phih(\mathsf{x})\phih(\mathsf{x}')\rangle$. Some of the most relevant states in quantum field theory, such as the vacuum state of a free scalar quantum field, are zero-mean Gaussian states (also called \textit{quasifree}), and are therefore fully described by their two-point functions.   
If we consider $N$ modes of a quantum field in a Gaussian state, the corresponding reduced state of the system modes is also Gaussian, since it is entirely represented by the one and two-point correlators of the canonical variables, $\langle \hat\Xi^\alpha \rangle$ and $\langle \hat\Xi^\alpha \hat\Xi^\beta \rangle$, or, equivalently, by its vector of means and covariance matrix,
\begin{equation}
    \bm\xi_0 = \langle \bm{\hat\Xi} \rangle, \quad \bm\sigma = 2 \operatorname{Re}\langle (\bm{\hat\Xi} - \bm\xi_0) (\bm{\hat\Xi}-\bm\xi_0)^\intercal \rangle.
\end{equation}

When it comes to the analysis of entanglement, the simplification brought by the restriction to Gaussian states is two-fold. On the one hand, {a generic} Gaussian state is fully characterized by only $N(2N+3)$ parameters ---with the restriction $\bm\sigma \geq \ii \bm\Omega^{-1}$, which is implied by the fulfillment of the CCR--- 
as opposed to generic states which are drawn from an infinite-dimensional Hilbert space. On the other hand, bipartite entanglement turns out to be very simple to quantify for Gaussian bisymmetric states\footnote{These are bipartite Gaussian states that are invariant under internal permutations of modes within either side of the partition.} 
and bipartitions of one versus  $M<N$ modes, where the Peres-Horodecki separability criterion is not only a sufficient but also a necessary condition~\cite{Simon2000, Serafini2005}. This means that in these Gaussian scenarios, separable states are exactly those with a positive partial transpose (PPT), and therefore the negativity and the logarithmic negativity are \textit{faithful} entanglement monotones. 
Moreover, even for cases where logarithmic negativity and negativity are not faithful entanglement monotones, they are still useful to characterize distillable entanglement~\cite{VidalNegativity, entanglementmeasuresreview}. 

Consider a bipartition of the $N$ total field modes into two complementary sets A and B of $N_\textsc{a}$ and $N_\textsc{b}$ modes, respectively. If the system of $N$ modes is in a Gaussian state of covariance matrix $\bm\sigma$, then we define the covariance matrix of the partial transpose with respect to B, $\tilde{\bm\sigma}$, as the result of reversing the sign of the momenta associated with system B: 
\begin{equation}
\tilde{\bm\sigma} = (\openone_{\textsc a} \oplus T_\textsc{b}) \bm\sigma (\openone_{\textsc a} \oplus T_\textsc{b}), 
\end{equation}
where
\begin{equation}
T_\textsc{b} = \bigoplus_{j=1}^{N_\textsc{b}}\begin{pmatrix}
    1 & 0 \\
    0 & -1
\end{pmatrix}.
\end{equation}
Let $\{\tilde{\nu}_1,\hdots,\tilde{\nu}_{N}\}$ be the symplectic spectrum of $\tilde{\bm\sigma}$, given by the absolute value of the eigenvalues of $\tilde{\bm\sigma}^{\alpha\beta}\Omega_{\beta\gamma}$, 
then the logarithmic negativity is given by~\cite{VidalNegativity}
\begin{equation}\label{Eq: logarithmic negativity from PPT symplectic eigenvalues}
    E_{\mathcal{N}} = \sum_{j=1}^{N} \max\{0,-\log_2\tilde{\nu}_j\}.
\end{equation}
In particular, A and B are entangled if and only if $\tilde{\sigma}$ has at least one symplectic eigenvalue strictly below 1, i.e., if and only if the condition $\tilde{\bm\sigma} \geq \ii \bm\Omega^{-1}$ is violated.

\subsection{Bipartite entanglement of spacelike separated modes in flat spacetime}\label{Subsection: Bipartite entanglement of spacelike separated modes in flat spacetime}

Once a finite subset is selected from the infinitely many degrees of freedom held by a quantum field, studying the entanglement between them amounts to studying the entanglement between a finite set of quantum harmonic oscillators---which is particularly simple when these are in a Gaussian state, as summarized in the previous subsection. Clearly, the entanglement between a finite number of modes of the quantum field cannot represent the full entanglement structure of the quantum field theory, but it captures some of it. For instance, considering two disjoint subregions, $A$ and $B$, of a spacelike Cauchy hypersurface $\Sigma$, one can attempt to study the bipartite entanglement between the regions by constructing two sets of modes such that their associated smearing functions have their supports restricted to $A$ and $B$, respectively. 
Examples of such modes are obtained by considering field and momentum operators of the form
\begin{align}\label{Eq: field and momentum modes fixed slice}
    \hat\Phi(F) & \coloneqq \int_{\Sigma} \dd\Sigma\, F \,\phih, \\
    \hat\Pi(F) & \coloneqq \int_{\Sigma} \dd\Sigma\, F \, n^a \nabla_a\phih, 
\end{align}
where $F$ is normalized to satisfy
\begin{equation}\label{Eq: F normalized}
    \int_\Sigma \dd\Sigma\, F^2 = 1,
\end{equation}
so that $[\hat\Phi(F),\hat\Pi(F)]=\ii\hat{\openone}$, hence $(\hat\Phi(F),\hat\Pi(F))$ defines a mode of the field. Then, if $\operatorname{supp} F \subset A$ (resp. $B$), the corresponding mode defined by $F$ is ``supported in'' $A$ (resp. $B$), in the sense that it belongs to the local algebra of $A$ (resp. $B$), or, equivalently, to the local algebra of the domain of dependence of $A$ (resp. $B$). 

The extent to which this strategy can be used to analyze the entanglement structure of quantum field theories is, however, limited by the ability of these kinds of modes to account for the entanglement present in the field. There is one reason why one might expect to be able to witness entanglement without having to fine-tune the choice of modes: the Reeh-Schlieder theorem (reviewed in section~\ref{sec:entInQFT}) guarantees that, under certain conditions on the state of the field $\ket{\Omega}$, the subregions $A$ and $B$ are entangled enough that the effect of any operator of the algebra of $A$ on $\ket{\Omega}$ can be approximated with arbitrary precision by the effect of operators of the algebra of $B$, and vice versa. Strictly speaking, this result only implies that, given a mode of region $A$, there exists a mode of region $B$ with which it is entangled. However, when thinking about it intuitively, the Reeh-Schlieder theorem could lead to the belief that entanglement is ``ubiquitous'' in QFT, i.e., that we should be able to witness entanglement between virtually \textit{any} pair of modes of regions $A$ and $B$ that we choose to define. 

This intuition was tested---and disproven---in~\cite{ubiquitous} for the simple case of the vacuum of a massless real scalar quantum field in a $(1+D)$-dimensional Minkowski spacetime, a state that fulfills the conditions of the Reeh-Schlieder theorem, and is additionally a Gaussian state. Specifically, for two spacelike separated spherical regions of the same radius $R$ belonging to the same time slice, centered in $\bm x_\textsc{a}$ and $\bm x_\textsc{b}$, respectively, we can consider modes of the form $(\hat\Phi(F_\textsc{i}),\hat\Pi(F_\textsc{i}))$, where \mbox{$F_\textsc{i}(\bm x)=F(\bm x - \bm x_\textsc{i})$}, for $\text{I}\in\{\text{A,B}\}$, and $F$ belongs to the family of smearing functions\footnote{The reader may rightfully point out that the functions in this family are not smooth. However, as shown in appendix B of~\cite{ubiquitous}, for any $\delta>0$, $F^{(\delta)} \in \dot{H}^{-1/2}(\mathbb{R}^D) \cap \dot{H}^{1/2}(\mathbb{R}^D) \subset L^2(\mathbb{R}^D)$, where $\dot{H}^{\pm 1/2}(\mathbb{R}^D)$ are homogeneous Sobolev spaces. This guarantees that the associated field and momentum operators are well defined, and that they can be arbitrarily approximated by smooth functions (since $\mathcal{C}_0^\infty(\mathbb{R}^D)$ is dense in $L^2(\mathbb{R}^D)$).}
\begin{equation}\label{Eq: smearing family}
    F^{(\delta)}(\bm x) = A_\delta \bigg(  1 - \frac{|\bm x|^2}{R^2} \bigg)^{\!\delta} \, \Theta\bigg( 1- \frac{|\bm x|}{R} \bigg)
\end{equation}
for $\delta\geq 1$, $\Theta$ is the Heaviside function, and~\cite{ubiquitous}
\begin{equation}\label{Eq: A_delta}
A_\delta = \sqrt{\frac{\Gamma(1+2\delta+D/2)}{\pi^{D/2} R^D \Gamma(1+2\delta)}}
\end{equation}
is the normalization constant that ensures that $F^{(\delta)}$ satisfies Eq.~\eqref{Eq: F normalized}. It was shown in~\cite{ubiquitous} that, for $|\bm x_\textsc{a} - \bm x_\textsc{b}| \geq 2R$, i.e., as long as the two regions do not overlap,
\begin{equation}
    E_{\mathcal{N}} = 0, \;\; \forall \delta\geq 1, \, D\geq 2,
\end{equation}
where $E_{\mathcal{N}}$ is the logarithmic negativity between the two modes of regions $A$ and $B$. Notice that in this case the logarithmic negativity is a faithful entanglement monotone, and hence $E_\mathcal{N}=0$ means that the joint state of the selected modes of regions $A$ and $B$ is separable. It is worth remarking that the family of smearing functions used to obtain this result are not pathological: they are compactly supported in the sphere of radius $R$, of differentiability class $\delta\geq 1$, spherically symmetric, and they peak at the origin. These are functions similar to what one would consider suitable to model, for instance, the spatial smearing of a compactly supported particle detector. Moreover, in~\cite{ubiquitous}, other simple spherically symmetric smearing functions{, including some functions  changing sign within their region of support,} are studied, observing the same result. 

The findings of~\cite{ubiquitous} are not in contradiction with the Reeh-Schlieder theorem: as stressed above, the theorem only guarantees that given a mode supported in $A$, there exists a mode supported in $B$ to which the former is entangled with. It does not specify how many modes in region B are entangled with the given mode in region A, nor does it describe the complexity of their spatial distribution.
In fact, the two modes that maximize their entanglement while being respectively supported in the spherical regions $A$ and $B$ are described in the lattice in~\cite{NatalieUVIR}, 
for $D=1,2,3$, showing that the shapes of these modes are not similar to the ones considered in~\cite{ubiquitous}, since they are not spherically symmetric.
The takeaway message from \cite{NatalieUVIR,ubiquitous} is that entanglement between individual modes is elusive unless the spatial profile of the modes under consideration is fine-tuned.  

\section{Entanglement Harvesting}\label{sec:entHarv}

A different approach to study the entanglement structure of a quantum field is given by the relativistic quantum information protocol that has become known as \textit{entanglement harvesting}. The protocol was first considered in~\cite{Valentini1991,Reznik2003,reznik1,reznik2}, and has been further consolidated in the more recent studies~\cite{Pozas-Kerstjens:2015,Pozas2016,HarvestingQueNemLouko,nogo,ericksonWhen}. It provides a way of extracting entanglement between two localized regions of a quantum field  by using probes that couple to the field in the two regions. One would expect that a detailed analysis of the origin of the entanglement acquired by the probes would provide insights into the available entanglement in the regions to which the detectors couple. The goals of this section are to review the commonly employed description of entanglement harvesting and to analyze the necessary conditions that enable probes to extract entanglement from a quantum field.

\subsection{Particle detectors coupled to a field}\label{sub:PDreview}

The probes used in the protocol of entanglement harvesting are typically modelled as particle detectors, which are localized quantum systems that couple to a quantum field. These were first considered by Unruh~\cite{Unruh1976} and DeWitt~\cite{DeWitt}, which led to these models also being known as Unruh-DeWitt (UDW) detectors. Since their conception, particle detector models have been considered for many different applications. Besides their use as a simple model in theoretical studies, the UDW model has also been shown to capture the relevant physics of the light-matter interaction in quantum optics \cite{eduardoOld,Pozas2016,richard} and other similar experimentally accessible setups~\cite{Sabinprl,Sabin2,SwitchQEDUpTheLadder,tunableCouplingTowardsHarvesting,cisco2023harvesting}, thus making it a rather versatile tool in several lines of research in relativistic quantum information. 

In this subsection, we will briefly review the coupling of two-level UDW detectors to a real scalar quantum field in 3+1 dimensional Minkowski spacetime. In this context, a particle detector's internal quantum degree of freedom is defined in a Hilbert space $\mathcal{H}_\tc{d}\cong \mathbb{C}^2$. The detector is assumed to undergo an inertial trajectory in spacetime, which can be written as $\mf z(t) = (t,\bm x_0)$, where $(t,\bm x)$ is an inertial coordinate system. Its internal dynamics are determined by the free Hamiltonian
\begin{equation}
    \hat{H}_\tc{d} = \Omega\, \hat{\sigma}^+ \hat{\sigma}^-,
\end{equation}
where $\Omega>0$ represents the energy gap between the qubit's ground and excited states. For convenience, we denote the ground and excited states by $\ket{g}$ and $\ket{e}$, respectively, so that $\hat{\sigma}^- \ket{e} = \ket{g}$ and $\hat{\sigma}^- \ket{g} = 0$, with $\hat{\sigma}^+ = (\hat{\sigma}^-)^\dagger$. 

The interaction with the quantum field is assumed to happen in a localized region of spacetime, determined by the spacetime smearing function $\Lambda(\mf x)$, which is assumed to be compactly supported\footnote{This assumption is often relaxed to allow functions that do not have compact support but are strongly localized around the detector's trajectory.} in both space and time. We further assume that $\Lambda(t,\bm x)$ factors as \mbox{$\Lambda(t,\bm x) = \chi(t) F(\bm x)$}, where $\chi(t)$ is a switching function which controls the time duration of the interaction, and $F(\bm x)$ is a smearing function which controls the shape of the interaction in the $(t,\bm x)$ frame. The detector is assumed to couple linearly to the quantum field through its monopole moment, which can be written as \mbox{$\hat{\mu}(t) = e^{\ii \Omega t}\hat{\sigma}^+ + e^{- \ii \Omega t}\hat{\sigma}^-$} in the interaction picture. One can then write the Hamiltonian that describes the interaction of the detector and field as\footnote{Although this prescription may not at first sight look covariant, this Unruh-DeWitt coupling can indeed be prescribed in a covariant manner. See~\cite{eduardo,us} for details.}
\begin{equation}
    \hat{H}_I(t) = \lambda \chi(t) \hat{\mu}(t) \int \dd^3 \bm x \, F(\bm x) \hat{\phi}(t,\bm x),
\end{equation}
where $\lambda$ is a dimensionless coupling constant that controls the interaction strength and $\hat{\phi}(\mf x)$ denotes the Klein-Gordon field. At each instant of time, the detector couples to the field observable defined by the smearing function $F(\bm x)$,
\begin{equation}
    \hat{\Phi}(t,F) = \int \dd^3 \bm x F(\bm x) \hat{\phi}(t,\bm x).\label{eq:PhiSigmat}
\end{equation}

In order to compute the final state of the detector after interacting with the field, one considers an initial uncorrelated state for the detector-field system, $\hat{\rho}_{\tc{d},0} \otimes \hat{\rho}_\phi$, with $\hat{\rho}_{\tc{d},0}$ being the initial state for the detector, and $\hat{\rho}_\phi$ being the field's state before the interaction. The time evolution of the system can be implemented by the unitary
\begin{equation}\label{eq:UI}
    \hat{U}_I = \mathcal{T}\exp(- \ii \int \dd t \, \hat{H}_I(t)),
\end{equation}
where $\mathcal{T}$ denotes the time ordering operation.
The compact support of $\chi(t)$ makes it so that the integral takes place over a finite time. The final state of the detector after the interaction, $\hat{\rho}_\tc{d}$, can be obtained by tracing over the field's degrees of freedom,
\begin{equation}
    \hat{\rho}_\tc{d} =\tr_{\phi}(\hat{U}_I (\hat{\rho}_{\tc{d},0} \otimes \hat{\rho}_\phi)\hat{U}_I^\dagger).
\end{equation}

 To describe the entanglement harvesting protocol, we consider two  detectors labelled by $\tc{A}$ and $\tc{B}$ that undergo inertial trajectories $\mf x_\tc{a}(t) = (t,\bm x_\tc{a})$ and $\mf x_\tc{b}(t) = (t,\bm x_\tc{b})$ with energy gaps $\Omega_\tc{a}$ and $\Omega_\tc{b}$ and spacetime smearing functions $\Lambda_\tc{a}(\mf x) = \chi_\tc{a}(t)F_\tc{a}(\bm x)$ and \mbox{$\Lambda_\tc{b}(\mf x) = \chi_\tc{b}(t)F_\tc{b}(\bm x)$}. 
The interaction Hamiltonian of the two detectors with the field can then be written as
\begin{equation}\label{eq:HIab}
    \hat{H}_I(t) = \lambda \left(\chi_\tc{a}(t)\hat{\mu}_\tc{a}(t)\hat{{\Phi}}_{\tc{a}}(t) + \chi_\tc{b}(t)\hat{\mu}_\tc{b}(t)\hat{{\Phi}}_{\tc{b}}(t)\right),
\end{equation}
where we denote
\begin{align}
    \hat{{\Phi}}_\tc{a}(t) =  \hat{{\Phi}}(t,F_\tc{a}),  \quad \hat{{\Phi}}_\tc{b}(t) =  \hat{{\Phi}}(t,F_\tc{b}),
\end{align}
and $\hat{\Phi}(t,F)$ is defined in Eq.~\eqref{eq:PhiSigmat}.
If the initial state of the detectors-field system is $\hat{\rho}_{\tc{ab},0}\otimes \hat{\rho}_{\phi}$ with \mbox{$\hat{\rho}_{\tc{ab},0} = \ket{g_\tc{a}g_{\tc{b}}}\!\!\bra{g_\tc{a}g_{\tc{b}}}$} (both detectors are initially in their ground states) and $\hat{\rho}_\phi$ is a zero-mean Gaussian state of the field, we find that the final state of the detectors, at leading order in $\lambda$, is of the form
\begin{equation}\label{eq:rhoab}
    \hat{\rho}_\tc{ab} = \begin{pmatrix}
        1 - \mathcal{L}_\tc{aa} - \mathcal{L}_\tc{bb} & 0 & 0 & \mathcal{M}^*\\
        0 & \mathcal{L}_\tc{bb} & \mathcal{L}_\tc{ab}^* & 0 \\
        0 & \mathcal{L}_\tc{ab} & \mathcal{L}_\tc{aa} & 0\\
        \mathcal{M} & 0 & 0 & 0
    \end{pmatrix} + \mathcal{O}(\lambda^4)
\end{equation}
when written in the basis $\{\ket{g_\tc{a}g_\tc{b}},\ket{g_\tc{a}e_\tc{b}},\ket{e_\tc{a}g_\tc{b}},\ket{e_\tc{a}e_\tc{b}}\}$. The terms $\mathcal{L}_{\tc{ij}}$ and $\mathcal{M}$ are given by
\begin{align}
    \mathcal{L}_{\tc{ij}} &= \lambda^2\! \int \dd t \dd t' \chi_\tc{i}(t)\chi_\tc{j}(t')e^{- \ii(\Omega_\tc{i} t - \Omega_\tc{j} t')}\langle \hat{\Phi}_{\tc{i}}(t)\hat{\Phi}_{\tc{j}}(t')\rangle,\nonumber\\
    \mathcal{M} &= -\lambda^2 \! \int \dd t \dd t' \chi_\tc{a}(t)\chi_\tc{b}(t')e^{ \ii(\Omega_\tc{a} t + \Omega_\tc{b} t')}\langle \mathcal{T} \hat{\Phi}_{\tc{a}}(t)\hat{\Phi}_{\tc{b}}(t')\rangle.\label{eq:LM}
\end{align} 
The state $\hat{\rho}_\tc{ab}$ in~\eqref{eq:rhoab} encodes the  changes in the detectors state after interaction with the field for weak interactions in localized spacetime regions. The terms $\mathcal{L}_\tc{aa}$ and $\mathcal{L}_\tc{bb}$ correspond to the leading order excitation probabilities of detectors $\tc{A}$ and $\tc{B}$, respectively. These local terms contain information about the entanglement between the field and each detector, while the terms $\mathcal{L}_{\tc{ab}}$ and $\mathcal{M}$ describe correlations acquired  by the detectors.

A relevant question is whether it is possible that the detectors end up entangled with each other after their interactions with the field. For a bipartitite qubit system, one can use the negativity~\cite{VidalNegativity} to quantify the entanglement between the detectors, since for two qubits it is a faithful entanglement measure in this setup. To leading order in $\lambda$, the partial transpose of the density matrix in Eq.~\eqref{eq:rhoab} has only one potentially negative eigenvalue, given by
\begin{equation}
    E = -\sqrt{|\mathcal{M}|^2 - \left(\frac{\mathcal{L}_{\tc{aa}} - \mathcal{L}_\tc{bb}}{2}\right)^2} + \frac{\mathcal{L}_\tc{aa} + \mathcal{L}_\tc{bb}}{2}.
\end{equation}
The logarithmic\footnote{Notice that the logarithmic negativity $E_\mathcal{N}$  is proportional to the negativity $\mathcal{N}$ at leading order in perturbation theory: \mbox{$E_\mathcal{N}=\log_2(1+2\mathcal{N})= 2\mathcal{N}+\mathcal{O}(\lambda^4)$.}} negativity can then be written as
\begin{equation}\label{Eq: negativity of a pair of detectors}
    E_\mathcal{N}(\hat{\rho}_{\tc{ab}}) = \max\big(0,-2E\big)+\mathcal{O}(\lambda^4).
\end{equation}

In essence, the detectors can become entangled if the $\mathcal{M}$ term is sufficiently larger than the average between $\mathcal{L}_\tc{aa}$ and $\mathcal{L}_\tc{bb}$. The negativity at leading order is thus a competition between the individual excitation probabilities that act as local noise, and the $|\mathcal{M}|$ term, which involves an integral of the (time ordered) correlations between $\hat{\Phi}_\tc{a}(t)$ and $\hat{\Phi}_\tc{b}(t)$ smeared in time with $\chi_\tc{a}(t)$ and $\chi_\tc{b}(t)$, respectively.


\subsection{When can two detectors harvest entanglement from a quantum field?}\label{sub:whenHarv}
As discussed earlier,
the two detectors can end up entangled through their interaction with a field. There are two mechanisms that enable this entanglement: 1) the detectors can communicate through the field, 
and 2) the field might also be in a state that contains entanglement between the coupling regions, and the detectors may `harvest' part of that entanglement. 
It is therefore important that, when we analyze the protocol of entanglement harvesting, we make sure that the detectors become entangled due to extraction of entanglement previously present in the field, and not due to communication, therefore allowing the particle detectors to be used as tools to access and study the entanglement of the quantum field.

One way to ensure that there is no signalling between the detectors is to consider spacelike separated interaction regions, so that the supports of $\Lambda_\tc{a}(\mf x)$ and $\Lambda_\tc{b}(\mf x)$ are causally disconnected. Indeed, it is possible to find many different scenarios where spacelike separated detectors\footnote{Note that in most examples of entanglement harvesting the spacetime smearing functions considered are not compactly supported, but rather they are strongly localized around  a finite spacetime region (e.g., Gaussian smearing). One can then quantify the effective communication between the detectors and ensure that it is negligible compared to the entanglement extracted from the field using techniques discussed in, e.g.,~\cite{ericksonWhen}.} can extract entanglement from a quantum field (see e.g.~\cite{Valentini1991,Reznik2003,reznik1,reznik2,Salton:2014jaa,Ng2014,mutualInfoBH,freefall,Pozas-Kerstjens:2015,Pozas2016,HarvestingSuperposed,Henderson2019,bandlimitedHarv2020,ampEntBH2020,carol,boris,ericksonWhen,threeHarvesting2022,twist2022,cisco2023harvesting,SchwarzchildHarvestingWellDone}). 
On the other hand, there are also many regimes where spacelike entanglement cannot be harvested. In particular, there are families of such scenarios where the harvesting of entanglement is forbidden that are covered by a \textit{no-go theorem}~\cite{nogo}, which we briefly summarize below.


When two detectors interact with the field in spacelike separated regions, we can factor the unitary time evolution $\hat{U}_I$ as 
\begin{equation}
    \hat{U}_I  = \hat{U}_\tc{a}\hat{U}_\tc{b}= \hat{U}_\tc{b}\hat{U}_\tc{a},
\end{equation}
where $\hat{U}_\tc{a}$ acts only on detector A and  on the field, and $\hat{U}_\tc{b}$ only acts on B and on the field. We can factor $\hat{U}_I$ as such because field operators smeared against the interaction regions A and B commute (as the supports of $\Lambda_\tc{a}$ and $\Lambda_\tc{b}$ are spacelike separated and the field satisfies the microcausality condition), and because observables that act on detector A commute with observables of detector B. Notice that because the unitary $\hat{U}_I$ factors as the product of two unitaries that act on systems A and B separately, $\hat{U}_I$ is unable to directly couple A and B. However, the field degrees of freedom supported in the interaction regions A and B can be entangled, which might allow entanglement to be exchanged between the field in regions A and B and the detectors. This is akin to an entanglement swap operation, where entanglement between the field degrees of freedom is transferred to entanglement between the detectors.



The no-go theorem in~\cite{nogo} points out specific cases where the unitary time evolution prescribed by the interaction with the field can be written as a simple-generated unitary. That is, when $\hat{U}_\tc{a} = e^{- \ii \hat{m}_\tc{a}\otimes \hat{X}_\tc{a}}$ or $\hat{U}_\tc{b} = e^{- \ii \hat{m}_\tc{b}\otimes \hat{X}_\tc{b}}$ for operators $\hat{m}_{\tc{a}}$ and $\hat{m}_{\tc{b}}$ that act in the respective detectors Hilbert spaces and field observables $\hat{X}_\tc{a}$ and $\hat{X}_\tc{b}$ localized in each detector's coupling regions. In this case it is possible to show that at least one of the commuting quantum channels implemented in each detector is an entanglement breaking channel, which implies that the two detectors will end up in  a separable state after their interaction with the field and thus entanglement harvesting is not possible.

There are two notable cases where $\hat{U}_\tc{a}$ and $\hat{U}_\tc{b}$ are simple-generated unitaries: the case of gapless detectors and the case of delta-coupled detectors (i.e., detectors whose switching function is a Dirac delta). 
In the case of gapless detectors we have $\Omega_\tc{a} = \Omega_\tc{b} = 0$ so that the free evolution of the detectors monopole moments is trivial: $\hat{\mu}_\tc{i}(t) = \hat{\mu}_\tc{i}$. In this case one can check that $[\hat{H}_I(t),\hat{H}_I(t')] \propto \hat{\openone}$, so that the unitary time evolution operator can be computed using the Magnus expansion~\cite{HarvestingQueNemLouko,Landulfo}:
\begin{equation}
    \hat{U}_\tc{a} = e^{\ii\varphi_\textsc{a}}\,e^{- \ii \lambda \hat{\mu}_\tc{a} \hat{\phi}(\Lambda_\tc{a})}, \quad\quad 
    \hat{U}_\tc{b} = e^{\ii\varphi_\textsc{b}}\,e^{- \ii \lambda \hat{\mu}_\tc{b} \hat{\phi}(\Lambda_\tc{b}) },
\end{equation}
for some phases $\varphi_\textsc{a},\varphi_\textsc{b}$, where we use the notation defined in~\eqref{eq:spacetimesmearedfield} for fields smeared in the spacetime support of the interaction regions. In this case, the unitaries $\hat{U}_\tc{a}$ and $\hat{U}_\tc{b}$  are simple-generated unitaries thus implementing an entanglement breaking channel, as explicitly discussed in~\cite{nogo}.

The other important case in which the detectors cannot harvest spacelike entanglement is when the interaction region with the field is contained in a spacelike surface~\cite{HarvestingQueNemLouko,nogo}, commonly called the delta-coupled case. In this limit, the spacetime smearing functions can be written as $\Lambda_\tc{i}(\mf x) = \eta \delta(t-t_\tc{i}) F_\tc{i}(\bm x)$, where $t_\tc{i}$ are the times at which the (sudden) couplings happen. $\eta$ is a parameter with dimensions of time, and $F_\tc{i}(\bm x)$ is a smearing function that defines the spatial profile of the interaction regions. In this scenario, $F_\tc{i}(\bm x)$ defines the field observables to which each detector couples to. We then have 

\begin{equation}
    \hat{U}_\tc{a} = e^{- \ii \lambda \eta \hat{\mu}_\tc{a}(t_\tc{a}) \hat{\phi}_\tc{a}(t_\tc{a})}, \quad \hat{U}_\tc{b} = e^{- \ii \lambda \eta \hat{\mu}_\tc{b}(t_\tc{b}) \hat{\phi}_\tc{b}(t_\tc{b})},
\end{equation}
which are simple-generated so that the same logic applies, showing that spacelike entanglement harvesting cannot take place using delta-coupled detectors. 
Notice, however, that it is possible to harvest entanglement if the detectors' coupling is given by \textit{multiple} sudden interactions, that is, when the coupling is described by a linear combination of a sufficiently large number of terms, each represented by a delta coupling~\cite{JoseEdu2024Nonperturbative}.

\section{The paradox of entanglement harvesting}\label{sec:paradox}


In previous sections, we have seen that, in a \mbox{$(1+3)$-dimensional} Minkowski spacetime:
1) Two field modes defined by smearing the field and conjugate momentum with non-overlapping, spherically symmetric functions of compact support are generally not entangled. In particular, they are unentangled for the choice of smearing functions $F^{(\delta)}(\bm x)$ given in \eqref{Eq: smearing family}; and,  2) Two particle detectors interacting with the field in causally disconnected regions can become entangled and, specifically, they do become entangled if the spatial profile of the detectors is chosen to be described by the functions $F^{(\delta)}(\bm x)$, as we will see later.

These two results appear paradoxical because choosing the spatial support of each detector to be described by $F^{(\delta)}(\bm x)$ means the detectors couple to the field degrees of freedom obtained by smearing the field operators with the function $F^{(\delta)}(\bm x)$. Given that such field modes are not entangled to begin with, where does the entanglement harvested by particle detectors come from?

It is crucial to recognize that the formulation of this tension rests on the observation that a detector is able to interact with infinitely many field degrees of freedom. For Unruh-DeWitt detectors, the field modes the detector couples to are obtained by smearing the field operator {\em at each instant} with the function $F(\bm x)$  describing the shape of the interaction. 
In other words, a detector interacts with a one-parameter family of distinct field modes $(\hat\Phi(t,F),\hat\Pi(t,F))$ throughout its time evolution. 
Notice that these different field modes are not generically independent: since the supports of $F$ at $\Sigma_t$ and at $\Sigma_{t'}$ are in causal contact, $(\hat\Phi(t,F),\hat\Pi(t,F))$ and $(\hat\Phi(t',F),\hat\Pi(t',F))$ will (in general) not commute\footnote{ However, as shown explicitly below, one can distill from these modes a family of (canonical) independent modes using a symplectic version of the Gram-Schmidt orthonormalization algorithm.}.



The picture that that detectors are indeed coupling to infinitely many modes as they move through spacetime is supported mathematically by the analysis in \cite{JoseEdu2024Nonperturbative}, which demonstrates that the evolution undergone by one or more detectors can be approximated with arbitrary precision by replacing the original continuous switching function with a ``train of delta couplings'', i.e., using $\chi(t)=\sum_i^N c_i, \delta(t-t_i)$, with $c_i=\chi(t_i)$. 
In summary,~\cite{JoseEdu2024Nonperturbative} shows that 
final state of the detectors after the many delta-coupling interactions converges to the result obtained with a smooth $\chi(t)$ if $N$ is chosen sufficiently large. The delta coupling can be thought of as describing instantaneous couplings to a single mode of the field.

One natural place to look for the field entanglement that the detectors harvest is in the potential entanglement between field modes that detectors A and B interact with at {\em different times}. In other words, even though the field modes $(\hat\Phi(t,F_A),\hat\Pi(t,F_A))$ and $(\hat\Phi(t,F_B),\hat\Pi(t,F_B))$ are not entangled, it may be possible that $(\hat\Phi(t,F_A),\hat\Pi(t,F_A))$ and $(\hat\Phi(t',F_B),\hat\Pi(t',F_B))$ are entangled for some choices of $t$ and $t'$ within the interaction interval. 
This, however, turns out to \textit{not} be the case, as can be checked, using the same techniques employed in~\cite{ubiquitous} and revisited in section~\ref{sec:modesQFT}.

At this stage one could even wonder whether the internally non-relativistic nature of the detector degrees of freedom may be to blame for the acquired entanglement, and hence cast doubts on the protocol of entanglement harvesting itself (see the discussions in~\cite{max,maxReply}). However, this was proven not to be the case. First, it is well-known that particle detectors do not introduce spurious causality violations in practically all regimes where entanglement harvesting protocols were considered~\cite{eduardoOld,us,PipoFTL,us2}. More importantly, there exist implementations of the protocol with fully relativistic detectors modelled by localized relativistic QFTs~\cite{fullHarvesting2024}, which actually show that the entanglement harvesting with particle detectors lower-bounds the entanglement that fully relativistic probes can harvest. This dispels any suspicion that the detector models are to blame for the apparent tension between the two results.

These observations bring us to conjecture that the origin of the entanglement harvested is actually distributed among many modes. That is, even if no pair $(\hat\Phi(t,F_A),\hat\Pi(t,F_A)),(\hat\Phi(t',F_B),\hat\Pi(t',F_B))$ is entangled, it is still possible that multimode subsystems, made of $N_A$ modes which couple to detector A and $N_B$ modes coupled to detector B, contain entanglement. This is a natural conjecture, since it is the only place left where entanglement could be located. The goal of the next section is to confirm that this is the case, hence identifying the source of the entanglement harvested by the two detectors as having a genuine multimode origin.

\section{Solving the tension: Multimode entanglement in entanglement harvesting}\label{sec:results}

{While evaluating the presence of entanglement in a finite-dimensional Gaussian system is a relatively easy task, searching for multimode entanglement between the continuous family of modes the two detectors interacts with is prohibitive. We will bypass this difficulty by restricting to finite sets within these families. That is, we will search for multimode entanglement between $N$ of the modes each detector is coupled to. While this makes calculations feasible if $N$ is not too large, it has an obvious limitation: the absence of entanglement between such finite-dimensional subfamilies would leave our investigations inconclusive, since one could not rule out that multimode entanglement will show up for a more numerous family of modes (i.e., for larger $N$, or continuous families of modes). Luckily, we do not need to go to such extremes: 
we will be able to demonstrate the presence of multimode entanglement before reaching the limit of our computational capabilities. 

As mentioned above, the strategy of restricting to a subfamily of the field modes that each detector couples to is further justified by the analysis in \cite{JoseEdu2024Nonperturbative}, which shows that replacing the coupling between detector and field by a train of $N$ instantaneously interactions results in an excellent approximation, as long as $N$ is large enough compared to the specific parameters that define the detector interaction.  These instantaneous interactions effectively single out a set of $N$ field modes. The intuition behind this result is that detectors have a finite time resolution, mostly determined by their energy gap $\Omega$, and are unable to temporally resolve the interaction with the field beyond a minimum time step, so that a finite number of modes is enough to approximate the interaction. 





Our strategy is summarized as follows: in subsections~\ref{Subsection: Entanglement harvesting: the modes that the detectors see} and~\ref{Subsection: Representation of mode operators on a common time slice}, we identify a finite subset of the field modes an inertial Unruh-DeWitt detector interacts with. Since these modes are generally not independent of each other, the direct application of the techniques described in Section~\ref{sec:modesQFT} presents some difficulties. Hence, in section~\ref{Subsection: Symplectic Gram-Schmidt algorithm}, we begin by employing a symplectic version of the Gram-Schmidt orthonormalization algorithm to transform the original set of modes a detector couples to into a family of independent and canonically conjugated modes. After doing this for both detectors, we compute the covariance matrix including the two sets of symplectically orthogonal modes, and  from it  we compute the logarithmic negativity between the two sets of modes.

\subsection{Entanglement harvesting: the modes that the detectors see}\label{Subsection: Entanglement harvesting: the modes that the detectors see}


Let us first consider two identical Unruh-DeWitt particle detectors, A and B, at rest at positions $\bm x_\textsc{a}$ and $\bm x_\textsc{b}$ in a \mbox{(1 + 3)-dimensional} Minkowski spacetime, just as in section~\ref{sub:PDreview}. Both detectors are described as two-level quantum systems, whose excited and ground energy levels are separated by a gap $\Omega$. For their shapes, in the calculations shown in this section we will use smearing functions \mbox{$F_\textsc{i}(\bm x) = F^{(2)}(\bm x - \bm x_\textsc{i})$}, with $F^{(2)}$ given in Eq.~\eqref{Eq: smearing family} for $\delta=2$. These are spherically symmetric functions supported in a sphere of radius $R$. The two detectors couple to a massless real scalar quantum field via the interaction Hamiltonian given in Eq.~\eqref{eq:HIab}, and for the switching function we use the $C^2$ function
\begin{equation}\label{Eq: swirtching function example}
    \chi(t) = \begin{cases} 
        \displaystyle{\bigg( 1 - \frac{4 t^2}{T^2} \bigg)^{\!5/2}} &\!\!\!\!,\:\: t\in[-T/2,T/2]\\[4mm]
        \:\:\:\:\:\:\:\:\:0 &\!\!\!\!, \:\:\text{otherwise}
    \end{cases}
\end{equation}
which has compact support in the interval $[-T/2, T/2]$ and is symmetric around $t=0$, where it peaks. We choose $T = 40 R$, and $|\bm x_\textsc{a} - \bm x_\textsc{b}| = T + 2 R$. This is the closest that both detectors can be while remaining spacelike separated during the time interval they interact with the field. Under these conditions, we can use Eq.~\eqref{Eq: negativity of a pair of detectors} to compute the negativity between detectors A and B once the coupling between them and the field has been switched off. Fig.~\ref{fig:HarvestingDetectors} shows the result, for different choices of the energy gaps $\Omega$ of the detectors. This figure confirms that, for the setup chosen, the detectors are able to harvest entanglement.

\begin{figure}[h]
    \centering
    \includegraphics[width=0.49\textwidth]{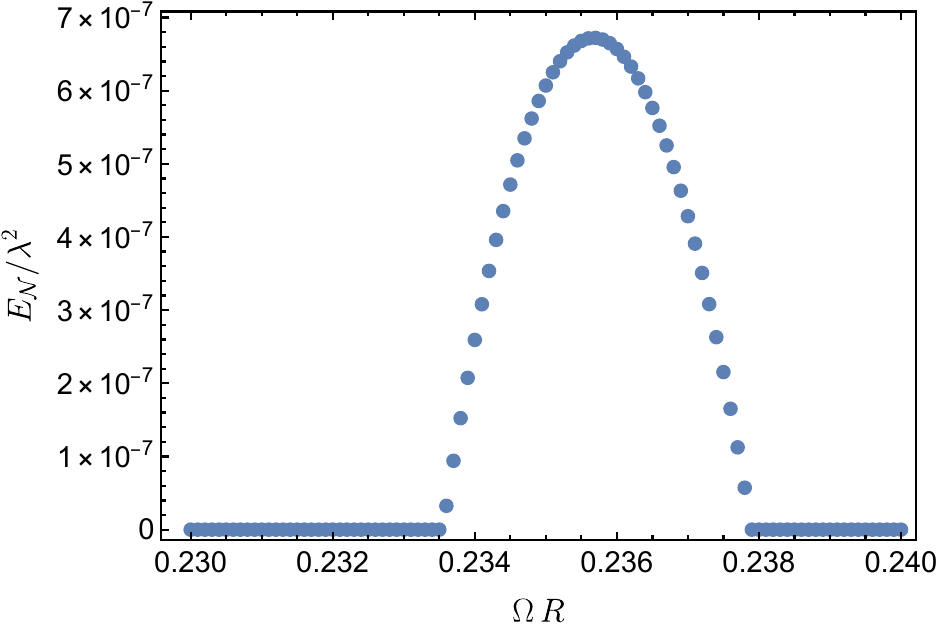}
    \caption{Negativity of two particle detectors evolving according to the switching and smearing functions given in section~\ref{Subsection: Entanglement harvesting: the modes that the detectors see} as a function of the energy gap $\Omega$   when $\Delta T = 40 R$. This figure shows that the entanglement the two detectors are able to harvest depends on their internal energy gap $\Omega$ and, for the chosen configuration, $\Omega R\approx 0.236$ is the optimal choice.}
    \label{fig:HarvestingDetectors}
\end{figure}

To search for multimode entanglement, we choose $N$ instants  within the interval of interaction
\begin{equation}
    t_i = -\frac{T}{2} + \frac{i}{N-1} T,
\end{equation}
for $i \in \{0,1,\hdots,N-1\}$. We restrict our attention to the family of field modes each detector is coupled with at instants $t_i$, defined by the $N$ pairs  \mbox{$(\hat\Phi_\textsc{i}^i=\hat\Phi(t_i,F_\textsc{i}),\hat\Pi_\textsc{i}^i=\hat\Pi(t_i,F_\textsc{i}))$}, for $\text{I}\in\{\text{A},\text{B}\}$.\footnote{Notice that the coupling of the detector at a given instant does not determine a unique momentum profile for the modes it couples to. However, the proper time of the trajectory of the detector introduces a preferred momentum, given by the derivative of $\hat\Phi(t_i,F_\textsc{i})$ with respect to the proper time of the detector.}

However, we cannot apply yet the tools spelled out in section~\ref{sec:modesQFT} to compute the entanglement between the two sets of modes. This is because the modes within each set (i.e., for $\text{I}=\text{A}$ or $\text{B}$) are not independent of each other, as the pairs $(\hat\Phi(t_i,F_\textsc{i}),\hat\Pi(t_i,F_\textsc{i}))$ do not necessarily commute for different values of  $i=1,\cdots,N$. 

However, as shown in the next subsections, one can bypass this difficulty by  distilling from the set of $N$ modes $(\hat\Phi(t_i,F_\textsc{i}),\hat\Pi(t_i,F_\textsc{i}))$ a set of independent (i.e., commuting) modes. We proceed in two steps. In the first step, we spell out the techniques we use to compute commutators of operators  $(\hat\Phi(t_i,F_\textsc{i}),\hat\Pi(t_i,F_\textsc{i}))$ defined at different times $t_i$. Once these commutators are known, we describe how to obtain a family of independent field modes.

\subsection{Representation of mode operators on a common time slice\label{Subsection: Representation of mode operators on a common time slice}}

Consider two out of the $N$ modes which interact with detector  $\tc{I}\in\{\tc{A},\tc{B}\}$, $(\hat\Phi(t_i,F_{\tc{i}}),\hat\Pi(t_i,F_{\tc{i}}))$ and $(\hat\Phi(t_j,F_{\tc{i}}),\hat\Pi(t_j,F_{\tc{i}}))$, with $t_i\neq t_j$. The commutators between these operators can be computed in two different, but equivalent ways. The most direct strategy is to use the covariant commutator $\big[\hat{\phi}(\mf x),\hat{\phi}(\mf x')\big] = \ii E(\mf x,\mf x')\hat{\mathds{1}}$, and its time derivatives $\big[\hat{\phi}(\mf x),\hat{\pi}(\mf x')\big] = \ii \partial_{t'} E(\mf x,\mf x')\hat{\mathds{1}}$ and $\big[\hat{\pi}(\mf x),\hat{\pi}(\mf x')\big] = \ii \partial_t\partial_{t'} E(\mf x,\mf x')\hat{\mathds{1}}$, 
and to smear each argument  with the functions $F_{\tc{i}}(\bm x) \delta(t-t_i)$ and $F_{\tc{i}}(\bm x') \delta(t'-t_j)$, respectively.

Equivalently, one can ``bring the field operators to a common time slice'' and use the canonical commutation relations. This strategy is based on the observation made at the end of section \ref{Subsection: Construction of field modes}, namely, smeared operators of the form
\begin{equation}\label{Eq: O varphi repeated}
\!\!\!\hat O(\varphi) = \!\!\int_{\Sigma_t}\!\!\!\dd\bm x\! \left(\sqrt{\mathsf{h}}\nabla_{\mathsf n}\varphi(t,\bm x) \, \hat{\phi}(t,\bm x) - \varphi(t,\bm x)\,\pih(t,\bm x)\right)\,
\end{equation}
with $\varphi(t,\bm x)$ a solution to the field equations, 
are independent of the slice ${\Sigma_t}$ chosen to evaluate the integral. Hence, by identifying the solution to the field equation $\varphi_{\Phi}(t,\bm x)$  having Cauchy data 
\begin{equation}\label{Eq: initial conditions for varphiPhi}
    \varphi_\Phi|_{t=t_j} = 0, \quad \partial_t\varphi_\Phi|_{t=t_j} = F_{\tc{i}},
\end{equation}
we can write the operator $\hat\Phi(t_j,F_\textsc{I})$ as an integral at time $t_i$:
\begin{equation}\label{Eq: Phi in t0}
\hat\Phi(t_j,F_{\tc{i}}) = \int_{\Sigma_{t_j}}\!\!\!\! \dd \bm x \, \Big[ \partial_t \varphi_\Phi(t_i,\bm x) \phih(t_i,\bm x) - \varphi_\Phi(t_i,\bm x) \pih(t_i,\bm x)   \Big]. 
\end{equation}
Similarly, we can write $\hat\Pi(t_j,F_{\tc{i}})$ as
\begin{equation}\label{Eq: Pi in t0}
    \hat\Pi(t_j,F_{\tc{i}})= \int_{\Sigma_{t_j}}\!\!\!\! \dd \bm x \, \Big[ \partial_t \varphi_\Pi(t_i,\bm x) \phih(t_i,\bm x) - \varphi_\Pi(t_i,\bm x) \pih(t_{i},\bm x)   \Big].
\end{equation}
where $\varphi_\Pi(t,\bm x)$ is the solution with initial data
\begin{equation}\label{Eq: initial conditions for varphiPi}
    \varphi_\Pi|_{t=t_j} = -F_{\tc{i}}, \quad \partial_t\varphi_{\Pi}|_{t=t_j} = 0\,.
\end{equation}
Writing all operators in terms of 
$\phih(t,\bm x)$ and $\pih(t,\bm x)$ evaluated at time $t_i$, makes it possible to use the equal-time canonical commutations relations \mbox{$[\phih(t_i,\bm x),\pih(t_i,\bm x)]=\ii\delta^{(3)}({\bm x}-{\bm x}')$}.

This second strategy for computing commutators of operators defined at different times requires finding the solutions $ \varphi_\Phi$ and $\varphi_\Pi$, defined by their intial data in \eqref{Eq: initial conditions for varphiPhi} and \eqref{Eq: initial conditions for varphiPi}, respectively. 
But this is a simple task, as we now summarize.

The solution to Klein-Gordon equation for a massless real scalar field in $(1+3)$-dimensional Minkowski spacetime at $t=t_0$ given initial data at $\tilde{t} \neq t_0$ is
\begin{align}
    \varphi(t_0,\bm x) = &\frac{1}{4\pi |\tilde{t}-t_0|} \bigg[ \int_{|\bm x - \bm y| = |t_0 -\tilde{t}|} \dd\bm y\, \partial_t\varphi(\tilde{t},\bm y)\\
    &\:\:\:\:\:\:\:\:\:\:\:\:\:\:\:\:\:\:\:\:\:\:\:+ \frac{\partial}{\partial t_0} \int_{|\bm x - \bm y| = |t_0 -\tilde{t}|} \dd\bm y\, \varphi(\tilde{t},\bm y) \bigg].\nonumber
\end{align}
Additionally, if the initial data is spherically symmetric, the solution will also display spherical symmetry. In that case, we have that the general solution, when evaluated at $t_0$, has the form
\begin{align}
    \varphi(t_0,r) & = \frac{1}{2r}\bigg[ u \,\bar\varphi(\tilde{t}, u) + v \, \bar\varphi(\tilde{t},v)   \bigg] \\
    & \phantom{====}+ \frac{1}{2r} \int_{\abs{u}}^{\abs{v}}\!\!\dd s\, s\, \partial_t \bar\varphi(\tilde{t}, s),
\end{align}
where $r = |\bm x|$, $u=r - (t_0 - \tilde{t})$, $v = r + (t_0 - \tilde{t})$, and we have defined
\begin{equation}\label{Eq: spherically symmetric solution}
    \bar\varphi(t,r) \coloneqq \frac{1}{4\pi r^2} \int_{|\bm x| = r} \dd\bm x\, \varphi(t,\bm x).
\end{equation}
In the particular case of the initial value problems for $\varphi_\Phi$ and $\varphi_\Pi$, where the initial conditions $\varphi(\tilde{t},\cdot)$ and $\partial_t\varphi(\tilde{t},\cdot)$ are given by Eqs.~\eqref{Eq: initial conditions for varphiPhi} and~\eqref{Eq: initial conditions for varphiPi}, respectively, we find that Eq.~\eqref{Eq: spherically symmetric solution} yields the simple analytic expressions
\begin{equation}
    \varphi_\Phi(t_0, r) = \frac{R^2}{12 r} \bigg[\bigg( 1-\frac{u^2}{R^2} \bigg)F^{(2)}(u)  - \bigg( 1-\frac{v^2}{R^2} \bigg)F^{(2)}(v) \bigg],
\end{equation}
and
\begin{equation}
    \varphi_\Pi(t_0, r) = -\frac{R}{2 r} \bigg[ \frac{v}{R}F^{(2)}(v) + \frac{u}{R}F^{(2)}(u) \bigg],
\end{equation}
with $F^{(2)}$ given by Eq.~\eqref{Eq: smearing family}.
%
By applying this procedure we can compute the commutator between the operators associated with the $N$ modes  detector \tc{I} interacts with.

Let \mbox{$\{(\hat\Phi^i_{\tc{i}},\hat\Pi_{\tc{i}}^i)\,:\,i = 0,1,\hdots,N-1\, , \, \tc{I}\in  \{\tc{A},\tc{B}\}\}$} be the original set of modes defined by the coupling of either of the detectors, whose operators can now be expressed as
\begin{align}
    \hat\Phi_{\tc{i}}^i & = \int \dd \bm x \Big[ p_{\Phi}^i(\bm x) \phih(t_0,\bm x) - q_{\Phi}^i(\bm x) \pih(t_0,\bm x)   \Big], \\
    \hat\Pi_{\tc{i}}^i & = \int \dd \bm x \Big[ p_{\Pi}^i(\bm x) \phih(t_0,\bm x,) - q_{\Pi}^i(\bm x) \pih(t_0,\bm x)   \Big],
\end{align}
where we have \mbox{$q_{\Phi/\Pi}^i(\bm x) = \varphi_{\Phi/\Pi}(t_0,\bm x)$} and \mbox{$p_{\Phi/\Pi}^i(\bm x) = \partial_t \varphi_{\Phi/\Pi}(t_0,\bm x)$}. 
Using  the equal-time commutation relations for $\phih(t_0,\bm x)$ and $\pih(t_0,\bm y)$, we obtain
\begin{align}\label{comm1}
[\hat\Phi_{\tc{i}}^i, \hat\Phi_{\tc{i}}^j] = \ii \alpha^{ij}\hat{\openone}, \quad [\hat\Pi_{\tc{i}}^i, \hat\Pi_{\tc{i}}^j] = \ii\, \beta^{ij} \hat{\openone},  \quad [\hat\Phi_{\tc{i}}^i, \hat\Pi_{\tc{i}}^j] = \ii \gamma^{ij}\hat{\openone},
\end{align}
where 
\begin{align}
\alpha^{ij} & = \int \dd\bm x\, \Big[ q_{\Phi}^i(\bm x) p_{\Phi}^j(\bm x) - p_{\Phi}^i(\bm x) q_{\Phi}^j(\bm x)  \Big], \\
\beta^{ij} & = \int \dd\bm x\, \Big[ q_{\Pi}^i(\bm x) p_{\Pi}^j(\bm x) - p_{\Pi}^i(\bm x) q_{\Pi}^j(\bm x)  \Big], \\ \label{gamma}
\gamma^{ij} & = \int \dd\bm x\, \Big[ q_{\Phi}^i(\bm x) p_{\Pi}^j(\bm x) - p_{\Phi}^i(\bm x) q_{\Pi}^j(\bm x)  \Big]\,.
\end{align}
The specific expressions of $\alpha^{ij}$, $\beta^{ij}$, and $\gamma^{ij}$ for our case are given in Appendix~\ref{Appendix: Commutation relations of the modes defined by the coupling of a detector}.

Using the expressions above, one can check that, whenever $N>\frac{T}{2R}+1$, the modes do not commute. The same happens for the modes detector B interacts with. On the other hand, because detectors A and B remain out of causal contact during the duration of the interaction, all field modes detector A interacts with commute with those interacting with B.

\savebox{\mybox}{\includegraphics[width=0.5\textwidth]{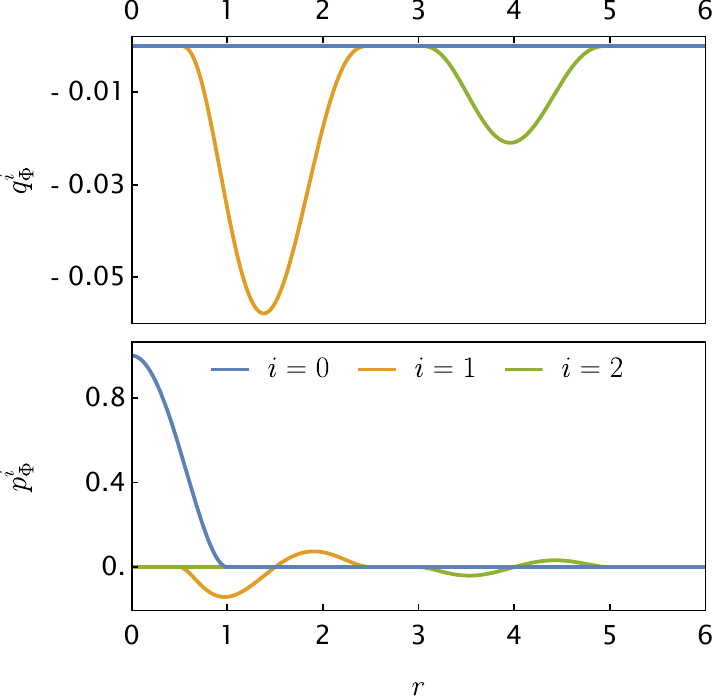}}
\begin{figure*}
\subfigure[]{\begin{minipage}{0.45\textwidth}
        \centering
        \vbox to \ht\mybox{%
            \vfill
            \hspace{-1cm}
            \resizebox{!}{0.85\ht\mybox}{

\begin{tikzpicture}
\draw[mathc2,fill=mathc2, fill opacity=0.1] (0, .5) -- (0.5,0) -- (2.5,0) --(1,1.5) --(0,1.5) --cycle ;

\draw[mathc3, fill=mathc3] (0,3.8) arc (-90:90:1cm and 0.2cm);

\draw[mathc3, fill=mathc3, fill opacity=0.1] (0,3) -- (3,0) -- (5,0) -- (1,4) --(0,4) --cycle; 
\draw[mathc3, fill=mathc3] (4,0) ellipse (1cm and 0.2cm);

\draw[draw=mathc1,fill=mathc1] (0,-0.2) arc(-90:90:1cm and 0.2cm); 
\draw[draw=mathc2,fill=mathc2] (0,1.3) arc(-90:90:1cm and 0.2cm); 

\draw[mathc2,fill=mathc2, fill opacity=0.9] (1.5,0) ellipse (1cm and 0.2cm);

\draw[->] (0,0) -- (0, 5) ; 
\draw[->] (0,0) -- (5.5, 0); 

\node[anchor=east] at (-0.15,5){$t$}; 
\node[anchor=west] at (5., -0.25) {$r$}; 
\draw (-0.1,1.5) -- (0.1,1.5) ; 
\draw(-0.1,4) -- (0.1,4); 
\node at (-0.25,4.) {$t_2$};
\node at (-0.25,0.) {$t_0$}; 
\node at (-0.25, 1.5) {$t_1$}; 

\end{tikzpicture}
}

            \vfill
        }
       
    \end{minipage}}~\subfigure[]{\begin{minipage}{0.45\textwidth}
        \centering
        \usebox{\mybox}
      
    \end{minipage}}
    \caption{Left panel: Depiction of the region of support of the backward evolution  from the time $t_i$ until $t_0$ of the phase space element $(0, F^{(2)}(\bm x))$, for two illustrative choices of $t_i$. Because $F^{(2)}(\bm x)$ is spherically symmetric, only the support in the radial coordinate is shown. Because the mode defined at $t_2$ does not overlap with the mode defined at $t_0$ when they are brought to the same time slice, these modes are guaranteed to be independent (commuting). In contrast, the mode defined at $t_1$ is not necessarily independent of that defined at $t_0$, because their supports overlap.  Right panel: Plot of the exact radial dependence of the phase space element $(q^i_{\hat{\Phi}}(\bm x), p^i_{\hat{\Phi}}(\bm x))$ obtained by propagating backwards in time $(0, F^{(2)}(\bm x))$ from $t_i$ to $t_0$ (same modes as in the left panel).}
    \label{fig:semi-artistic}
\end{figure*}
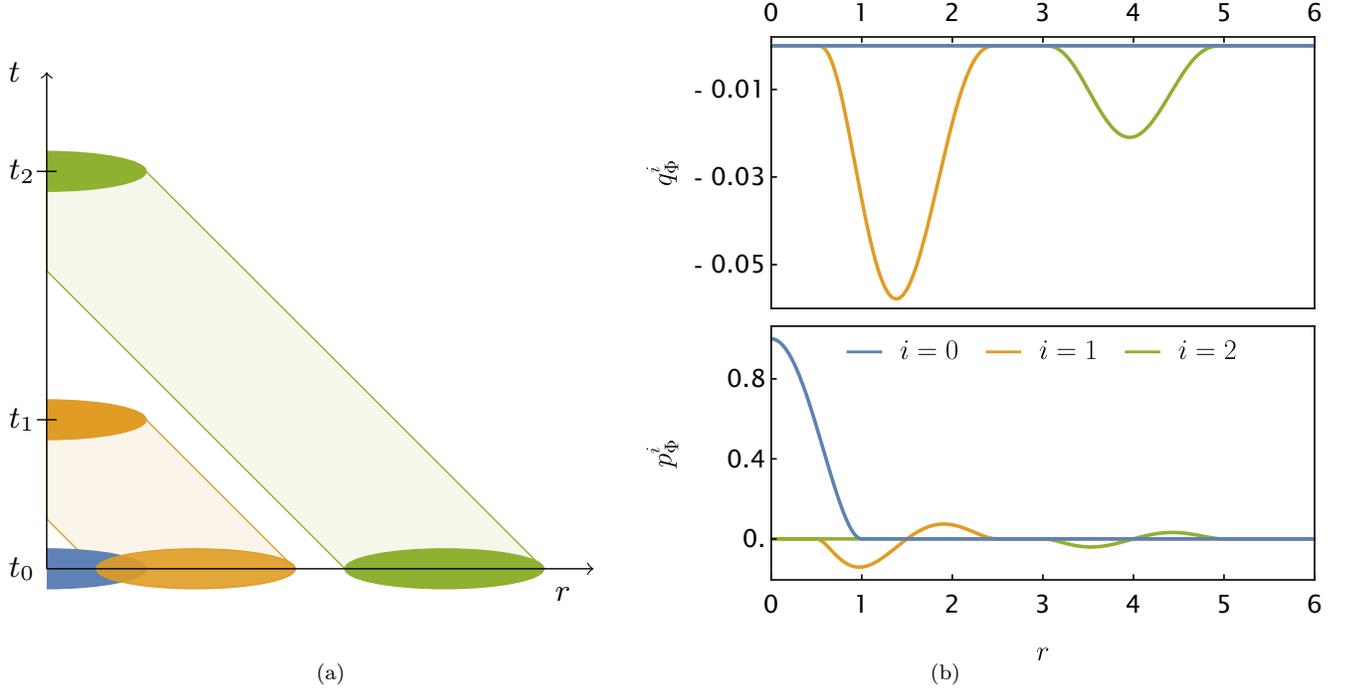

\subsection{Symplectic Gram-Schmidt algorithm\label{Subsection: Symplectic Gram-Schmidt algorithm}}

Once the commutators \eqref{comm1} are known, we 
can extract $N$ independent modes with the same associated joint Hilbert space as the original set.  This can be done by applying a symplectic analogue of the Gram-Schmidt orthonormalization algorithm, where the role of the inner product between vectors is played here by the commutator (see, e.g., Appendix A of~\cite{Natalie2}). 

Let us denote by $\{(\hat Q^i,\hat P^i)\,:\,i=0,1,\hdots,N-1\}$ the commuting set of modes we are looking for. They are obtained as follows: First, $\hat Q^1$ and $\hat P^1$  are given by
\begin{equation}
    \hat Q^1 = \frac{1}{\sqrt{\abs{\gamma^{11}}}}\,\hat\Phi^1, \quad \hat P^1 = \frac{\sqrt{\abs{\gamma^{11}}}}{\gamma^{11}}\,\hat\Pi^1, 
\end{equation}
where $\gamma^{ij}$ is given in Eq.~\eqref{gamma}. 

For the rest of the modes, $(\hat Q^k, \hat P^k)$ for $k \geq 2$, we first define an auxiliary pair of operators $(\hat X^k, \hat Y^k)$ as
\begin{align}\label{eq:provisional_SGS}
    \hat X^k & = \hat\Phi^k +\ii \sum_{j=1}^{k-1} \Big( [\hat Q^j, \hat\Phi^k] \hat P^j - [\hat P^j, \hat\Phi^k] \hat Q^j \Big), \\ \label{eq:provisional_SGS2}
    \hat Y^k & = \hat\Pi^k + \ii \sum_{j=1}^{k-1} \Big( [\hat Q^j, \hat\Pi^k] \hat P^j - [\hat P^j, \hat\Pi^k] \hat Q^j \Big),
\end{align}
where the commutators $[\hat Q^j, \hat\Phi^k], [\hat P^j, \hat\Phi^k], [\hat Q^j, \hat\Pi^k]$, and $[\hat P^j, \hat\Pi^k]$ appearing in this expressions are all proportional to $\hat{\openone}$, and the proportionality constant can be written in terms of the real coefficients $\alpha^{ij}, \beta^{ij}, \gamma^{ij}$. A straightforward  calculation yields
\begin{align}
    [\hat X^k, \hat Y^k] & = \gamma^{kk}\hat{\openone} + \ii \sum_{j=1}^{k-1} \Big( [\hat Q^j, \hat\Phi^k][\hat P^j, \hat\Pi^k] \nonumber\\
    & \phantom{=\;} - [\hat P^j, \hat\Phi^k][\hat Q^j, \hat\Pi^k]  \Big) \equiv \ii \bar\gamma^{kk}\hat{\openone}.
\end{align}
Finally, we  have
\begin{equation}
    \hat Q^k = \frac{1}{\sqrt{\abs{\bar\gamma^{kk}}}}\,\hat X^k, \quad \hat P^k = \frac{\sqrt{\abs{\bar\gamma^{kk}}}}{\bar\gamma^{kk}}\,\hat Y^k.
\end{equation}
The new modes $(\hat Q^i,\hat P^i)$, $(i=1,\cdots, N)$ satisfy $[\hat Q^i,\hat Q^j]=[\hat P^i,\hat P^j]= 0$, and $[\hat Q^i,\hat P^j]=\delta^{ij}$, i.e., they define a \emph{canonically commuting} set of $N$ modes that span the same subalgebra as the  non-commuting modes we started with. 

For later use, it is convenient to write the operators $\hat Q^i$ and $\hat P^i$ in terms of $\phih(t_0,\bm x)$ and $\pih(t_0,\bm x)$:
\begin{align}\label{Eq: independent modes in terms of phi and pi}
    \hat Q^i & = \int \dd \bm x \Big[ w_\textsc{q}^i(\bm x) \phih(t_0,\bm x) - z_\textsc{q}^i(\bm x) \pih(t_0\bm x)   \Big], \\
    \hat P^i & = \int \dd \bm x \Big[ w_\textsc{p}^i(\bm x) \phih(t_0,\bm x) - z_\textsc{p}^i(\bm x) \pih(t_0,\bm x)   \Big],
\end{align}
where $z_\textsc{q/p}^i,w_\textsc{q/p}^i$ are combinations of $\alpha^{ij}$, $\beta^{ij}$    and  $\gamma^{ij}$. 

 We have so far presented the algorithm to extract a set of independent modes in a way that is easier to understand. In practice, if one were to implement this algorithm numerically as we have presented it, they would encounter well-known numerical instabilities~\cite{bjorck_numerics_1994}. Although these numerical instabilities can be circumvented by using a modified version of the algorithm  (see, for instance,~\cite{salam_theoretical_2005}), this is not the strategy we use in this work. The reason is that the coefficients in Eqs.~\eqref{eq:provisional_SGS} and \eqref{eq:provisional_SGS2} must be computed with high precision, since numerical errors, even if small,  propagate and end up impacting the final result in a significant manner. 
We have bypassed these difficulties by analytically evaluating the coefficients in Eqs.~\eqref{eq:provisional_SGS} and \eqref{eq:provisional_SGS2}. This strategy has the advantage of producing an exact result. We use the software \emph{Mathematica} to manipulate the resulting analytical expressions, which are extremely long and have been omitted because they are not particularly illuminating. Despite being analytical, this calculation requires moderate resources, as the successive steps in the algorithm require a large number of calls to hypergeometric functions. (In the next subsection, we will use these analytical results to compute, this time numerically, the covariance matrix and logarithmic negativity).

\subsection{Multimode entanglement in spacetime\label{Subsection: Multimode entanglement in spacetime}}


Once we have $N$ independent modes associated with the the degrees of freedom to which each detector couples throughout its history, 
because the vacuum of a real scalar quantum field in Minkowski spacetime is a (zero-mean) Gaussian state, we can use the formalism described in section~\ref{Subsection: Gaussian states} to analyze the bipartite entanglement between the set of modes associated with detector A and the set of modes associated with detector B. The first step for this procedure is to calculate the covariance matrix
\begin{equation}
\bm\sigma = 2 \operatorname{Re}\langle\, \bm{\hat\Xi} \,\bm{\hat\Xi}^\intercal \rangle = \begin{pmatrix}
    \bm\sigma_\textsc{a} & \bm\eta \\
    \bm\eta^\intercal & \bm\sigma_\textsc{b}
\end{pmatrix},
\end{equation}
where we have used the vector of (ordered) canonical observables
\begin{equation}
    \bm{\hat\Xi} = (\hat Q_\textsc{a}^1,\hat P_\textsc{a}^1,\hdots,\hat Q_\textsc{a}^N,\hat P_\textsc{a}^N,\hat Q_\textsc{b}^1,\hat P_\textsc{b}^1,\hdots,\hat Q_\textsc{b}^N,\hat P_\textsc{b}^N)^\intercal.
\end{equation}
The translational symmetry between the modes defined by detector A and those defined by detector B yields the equality $\bm\sigma_\textsc{a} = \bm\sigma_\textsc{b} \equiv \bm\sigma_\textsc{d}$. In particular, since \mbox{$F_\textsc{i}(\bm x) = F^{(2)}(\bm x - \bm x_\textsc{i})$}, for $\text{I} \in \{\text{A},\text{B}\}$, and $F^{(2)}(\bm x)$ has spherical symmetry, we can define, for \mbox{$l \in \{0,\hdots,2N-1\}$},
{\begin{align}
    &\tilde{z}^l(|\bm k|)\coloneqq e^{\ii \bm k \cdot \bm x_\textsc{a}} \tilde{z}_\textsc{a}^l(\bm k) = e^{\ii \bm k \cdot \bm x_\textsc{b}} \tilde{z}_\textsc{b}^l(\bm k) , \\
    & \tilde{w}^l(|\bm k|)\coloneqq e^{\ii \bm k \cdot \bm x_\textsc{a}} \tilde{w}_\textsc{a}^l(\bm k) = e^{\ii \bm k \cdot \bm x_\textsc{b}} \tilde{w}_\textsc{b}^l(\bm k), 
\end{align} 
where the tilde denotes the Fourier transform\footnote{Here, we use the convention by which the Fourier transform is defined as
\begin{equation}
    \tilde{K}(\bm v) = \int \dd \bm u K(\bm u) \, e^{-\ii \bm u \cdot \bm v}.
\end{equation}} and the exponentials $e^{\ii \bm k \cdot \bm x_\textsc{a}}$ and $e^{\ii \bm k \cdot \bm x_\textsc{b}}$ are  needed because $F^{(2)}(\bm x - \bm x_\textsc{i})$ are centered at  $x_\textsc{i}$. Note also that, by allowing the superscript $l$ in ${z}_\textsc{b}^l$ to run from $0$ to $2N-1$, we have embedded in them the previously defined functions $z_\textsc{q a/b}$ and $z_\textsc{p a/b}$, with $i=1,\cdots,N$, in the following manner:
\begin{equation}
    z_\textsc{a/b}^{2i} = z_\textsc{q a/b}^i, \quad  z_\textsc{a/b}^{2i+1} = z_\textsc{p a/b}^i.
\end{equation} 
And similarly for ${w}_\textsc{a/b}^l$.} Notice that $z^l$ and $w^l$ are the functions that result from applying the same procedure needed to obtain ${z}_\textsc{a/b}^l$ and ${w}_\textsc{a/b}^l$, but with the modes defined from the spherically symmetric $F^{(2)}(\bm x)$, instead of the displaced $F_\textsc{a/b}(\bm x) = F^{(2)}(\bm x - \bm x_\textsc{a/b})$.

We can then compute the components of the covariance matrix as 
\begin{align}
    \sigma_\textsc{d}^{lm} & = \frac{1}{2\pi^2} \int_0^\infty\dd|\bm k|\,\Big[ |\bm k| \tilde{w}^l(|\bm k|) \tilde{w}^m(|\bm k|) \\
    & \phantom{==========} + |\bm k|^3 \tilde{z}^l(|\bm k|) \tilde{z}^m(|\bm k|)  \Big] \nonumber \\
    \eta^{lm} & = \frac{1}{2\pi^2} \int_0^\infty\dd|\bm k|\, \Big[ |\bm k| \tilde{w}^l(|\bm k|) \tilde{w}^m(|\bm k|) \\
    & \phantom{======} + |\bm k|^3 \tilde{z}^l(|\bm k|) \tilde{z}^m(|\bm k|)  \Big] \operatorname{sinc}(|\bm k||\bm x_\textsc{a} - \bm x_\textsc{b}|). \nonumber
\end{align}

{Although the final part of the calculation seems straightforward, several practical challenges arise. Namely, the complexity of the symplectically orthogonalized modes, the size of the covariance matrices for large $N$, and the high precision needed for determining the symplectic spectrum of the resulting covariance matrix. Specifically, the evaluation of the components of the covariance matrix requires integration, making the calculation numerically expensive when the number $N$ of modes is large.
We bypass some of these challenges by analytically computing the components of the covariance matrix $\sigma^{lm}_D$ as products of linear combinations of the original mode correlations (prior to symplectic orthogonalization). Additionally, we evaluate numerically the off-block diagonal components of the covariance matrix, $\eta^{lm}$. This semi-analytical approach eliminates numerical errors  in the evaluation of a significant number of components of the covariance matrix, enabling us to achieve sufficient precision in determining the symplectic spectrum.} 

{
Once the covariance matrix is obtained, we can follow the procedure given in section~\ref{Subsection: Gaussian states} to compute the logarithmic negativity, given by Eq.~\eqref{Eq: logarithmic negativity from PPT symplectic eigenvalues}, for different values of $N$. The results are shown in Fig.~\ref{fig: LogNeg different Ns}.

From the figure, we can appreciate the following points. First of all, we find that there exists multimode entanglement between the field modes that detectors A and B interact with. This is true even though no single mode coupled to detector A is entangled with any of the individual field modes coupled to detector B. This result confirms the existence of entanglement in the field modes the detectors have access to and showcases the genuine multimode nature of this entanglement. 

Secondly, we notice that there is a minimum number of modes that need to be included 
in our analysis to observe entanglement in the field. 
For instance, for the geometric configuration that we have used, $N$ needs to 
be at least\footnote{This may not be surprising for the reader of \textit{The Hitchhiker's Guide to the Galaxy. }} 42. This result is in consonance with the analysis done in \cite{JoseEdu2024Nonperturbative} by replacing the continuous switching function with a train of instantaneous delta-couplings. There, it was shown that a minimum number of instantaneous couplings are needed for the detectors to become entangled with each other. This observation can also be understood from the viewpoint of the \textit{no-go theorems} studied in \cite{nogo}. 

Third, within our computational capabilities, we observe that the logarithmic negativity acquires a plateau for large $N$, suggesting that the multimode entanglement within the field modes the detectors couple to may be finite and  well captured by analyzing a sufficiently large number of modes.

Finally, the values of the logarithmic negativity shown in Fig.~\ref{fig: LogNeg different Ns} are many orders of magnitude larger than the logarithmic negativity in the final state of the two detectors plotted in Fig.~\ref{fig:HarvestingDetectors}. 
 This result is not surprising, given that the detectors were analyzed only in the perturbative regime, where the harvested entanglement cannot be very large and the interaction parameters (observables, gap, coupling strength, etc.) cannot be fully optimized to maximize entanglement harvesting.
 }

\begin{figure}[h!]
    \centering
    \includegraphics[width=0.49\textwidth]{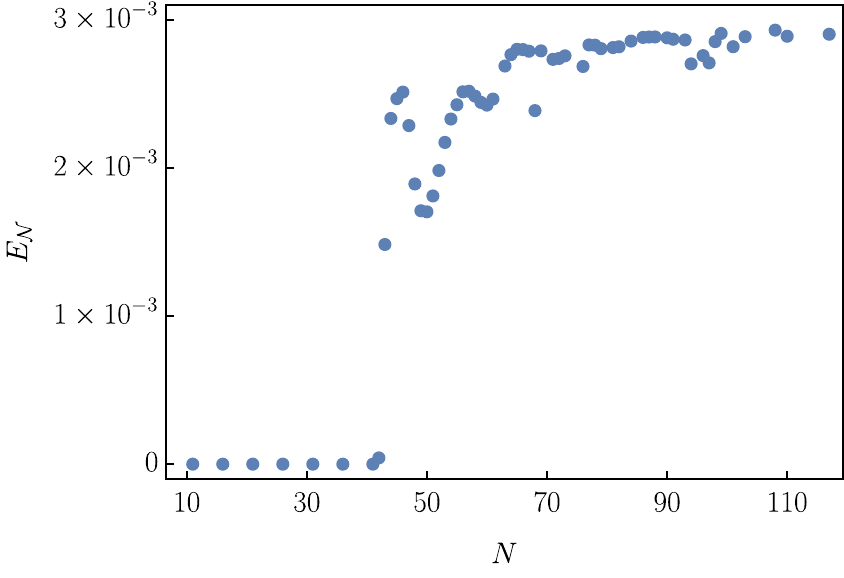}
    \caption{Logarithmic negativity quantifying the entanglement between the two sets of $N$ modes associated with detectors A and B, as a function $N$, when the separation between the centres of the regions A and B is $|\bm x_\textsc{a}-\bm x_\textsc{b}| = T + 2R$. }
    \label{fig: LogNeg different Ns}
\end{figure}

\section{Conclusion}\label{sec:discussions}




We explored 
the existence of multimode entanglement in quantum field theory, by analyzing entanglement between families of field degrees of freedom confined in finite regions of spacetime. In particular, we studied the case of a free, massless scalar field theory in Minkowski spacetime. This analysis is motivated by an apparent paradox involving the protocol of (spacelike) entanglement harvesting with Unruh-DeWitt detectors and the results of~\cite{ubiquitous} about the distribution of entanglement in QFT. Concretely, since the two detectors in the protocol interact with the field in spacelike separated regions of spacetime,we were able to identify a one-parameter family of field degrees of freedom that each detector interacts with. 
The apparent tension with the results of~\cite{ubiquitous} arises 
because one can find cases where none of the field degrees of freedom that detector A interacts with is entangled with any of the individual degrees of freedom coupled to detector B, raising questions about the origin of the entanglement harvested by the detectors. The answer to these questions is that the entanglement harvested by the detectors is of multimode nature.

Indeed, we have found that in all scenarios explored in this article, there exists multimode entanglement between sets of field degrees of freedom coupled to detectors A and B, even when there is no pairwise entanglement between individual modes. Our results make manifest that detectors can see beyond pairwise entanglement in the field, and can access entanglement which is distributed among many field modes. In other words, letting the two detectors interact with the field during a finite time interval allows the detectors to access field correlations which are distributed among many modes of the field. 

Our study also has implications for the investigation of the distribution of entanglement in QFT. The results of~\cite{Natalie1, NatalieUVIR, Natalie2, Natalie3} and \cite{ubiquitous} show that in (1+3) spacetime dimensions most pairs of field modes are unentangled, and in order to find pairs of entangled modes one needs to carefully select the modes under consideration. I.e., pair-wise entanglement is not as ubiquitous as one could have intuitively thought, becoming even less so in higher dimensions \cite{ubiquitous}.  
In contrast to this, our results indicate that multimode entanglement is quite ubiquitous. Namely, 
we have shown that even if one chooses sets of modes for which there is no pairwise entanglement across  regions, there still exists multimode entanglement. 
Furthermore, 
these results suggest that the entanglement content in the quantum field 
that is typically accessed by coupling the field to external probes, is predominantly multimode in nature.



\begin{acknowledgements}
    The authors would like to thank Natalie Klco for very helpful discussions. IA is supported by the Hearne Institute for Theoretical Physics, by the NSF grants PHY-2409402 and PHY-2110273, by the RCS program of Louisiana Boards of Regents through the grant LEQSF(2023-25)-RD-A-04, and by Perimeter Institute for Theoretical Physics. EMM acknowledges support through the Discovery Grant Program of the Natural Sciences and Engineering Research Council of Canada (NSERC). EMM also acknowledges support of his Ontario Early Researcher award. SNG acknowledges the financial support of the Research Council of Finland through the Finnish Quantum Flagship project (358878, UH). TRP acknowledges support from the Natural Sciences and Engineering Research Council of Canada (NSERC) through the Vanier Canada Graduate Scholarship. JPG acknowledges the support of a fellowship from “La Caixa” Foundation (ID 100010434, with fellowship code LCF/BQ/AA20/11820043). JPG and BSLT acknowledge support from the Mike and Ophelia Lazaridis Fellowship. PRM acknowledges the financial support from the Blaumann Foundation and the support provided by the Perimeter Institute for Theoretical Physics. Research at Perimeter Institute is supported in part by the Government of Canada through the Department of Innovation, Science and Industry Canada and by the Province of Ontario through the Ministry of Colleges and Universities. 
\end{acknowledgements}

\onecolumngrid
\appendix

\section{Commutation relations of the modes defined by the coupling of a detector}\label{Appendix: Commutation relations of the modes defined by the coupling of a detector}

In this appendix, we give analytic expressions for the coefficients 
 $\alpha^{ij}$, $\beta^{ij}$, and $\gamma^{ij}$, which were defined in Sec.~\ref{Subsection: Representation of mode operators on a common time slice} from the commutators $[\hat\Phi^i,\hat\Phi^j]$, $[\hat\Pi^i,\hat\Pi^j]$, and $[\hat\Phi^i,\hat\Pi^j]$, in Eq.~\eqref{comm1}. These can be computed directly, obtaining
\begin{align}
    \alpha^{ij}& = -2^{2 \delta +2}\pi  R^3\,A_{\delta}^2\, \Gamma (\delta +1)^2 \, \Delta t_{ij}\,\Theta \bigg(2-\left|\frac{\Delta t_{ij}}{R}\right|\bigg) \bigg[\frac{ \sin (\pi  \delta ) \Gamma (-2 \delta -2)}{\pi  } \left| \frac{\Delta t_{ij}}{R}\right| ^{2 \delta +1} \,\!\!\!\! _2F_1\bigg(\frac{1}{2},-\delta -1;\delta +\frac{3}{2};\frac{\Delta t_{ij}^2}{4 R^2}\bigg) \nonumber\\ 
    & \phantom{===} +\frac{\Gamma \big(\delta +\frac{1}{2}\big)}{4  \Gamma (\delta +1) \Gamma\big(2 \delta +\frac{5}{2}\big)}  \, _2F_1\bigg(\!\!-2 \delta -\frac{3}{2},-\delta ;\frac{1}{2}-\delta ;\frac{\Delta t_{ij}^2}{4 R^2}\bigg)\bigg],\\
    \beta^{ij} & = 2 \Delta t_{ij}\, R \,A_{\delta}^2\, \Theta(2R-|\Delta t_{ij}|) \bigg[2^{2 \delta +1} \sin (\pi  \delta ) \Gamma (-2 \delta ) \Gamma (\delta +1)^2 \left| \frac{\Delta t_{ij}}{R}\right| ^{2 \delta -1} \,\!\!\!\! _3F_2\bigg(\frac{1}{2},-\delta -1,\delta +2;\delta +\frac{1}{2},\delta +1;\frac{\Delta t_{ij}^2}{4 R^2}\bigg) \nonumber \\ 
    & \phantom{===} -\frac{3 \pi ^{3/2} \delta ^2 \Gamma (2 \delta -1)}{\Gamma \left(2 \delta +\frac{3}{2}\right)} \, _3F_2\bigg(\frac{5}{2},-2 \delta -\frac{1}{2},1-\delta ;\frac{3}{2},\frac{3}{2}-\delta ;\frac{\Delta t_{ij}^2}{4 R^2}\bigg)\bigg],\\
    \gamma^{ij} & = R^3 A_{\delta}^2\, \Theta(2R-|\Delta t_{ij}|)\bigg[\frac{\pi ^{3/2} \Gamma (2 \delta +1)}{\Gamma \big(2 \delta +\frac{5}{2}\big)} \, _3F_2\bigg(\frac{3}{2},-2 \delta -\frac{3}{2},-\delta ;\frac{1}{2},\frac{1}{2}-\delta ;\frac{\Delta t_{ij}^2}{4 R^2}\bigg) \nonumber \\
     & \phantom{===} -4^{\delta +1} \sin (\pi  \delta ) \Gamma (-2 \delta -1) \Gamma (\delta +1)^2 \left| \frac{\Delta t_{ij}}{R}\right| ^{2 \delta +1} \, \!\!\!\!  _3F_2\bigg(\frac{1}{2},-\delta -1,\delta +2;\delta +1,\delta +\frac{3}{2};\frac{\Delta t_{ij}^2}{4 R^2}\bigg)\bigg], \\
\end{align}
where $\Delta t_{ij} = t_i - t_j$, and $A_\delta$ is given in Eq.~\eqref{Eq: A_delta}. For illustration purposes, these expressions are plotted in Fig.~\ref{fig:sp_delta2_all} for the case $\delta =2$.

\begin{figure}[h]
    \centering
    \includegraphics[width=0.55\textwidth]{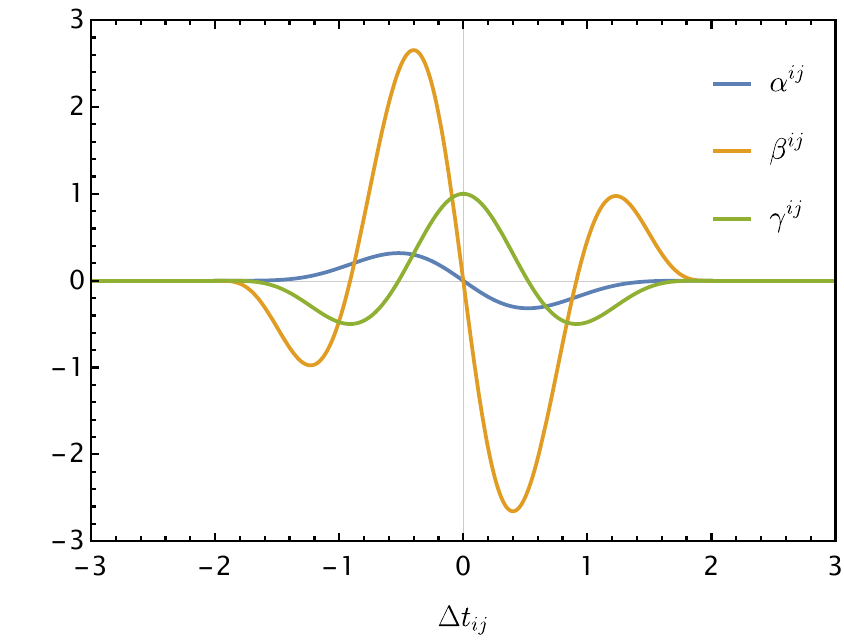}
    \caption{Symplectic product between field-field, momentum-momentum and field-momentum phase space elements defined from prescribing the initial data in Eq.~\eqref{comm1} as a function of the time difference $\Delta t_{ij} = t_i-t_j$.   }
    \label{fig:sp_delta2_all}
\end{figure}

\vspace{3cm}
\twocolumngrid

\bibliography{references.bib}

\end{document}